\documentclass[useAMS,usenatbib]{mn2e}

\usepackage{graphicx}
\usepackage{amssymb}

\newcommand{\apj}{ApJ}

\newcommand{\aap}{A\&A}

\newcommand{\aj}{AJ}
\newcommand{\mnras}{MNRAS}

\newcommand{\apjs}{ApJS}
\newcommand{\MC}{\multicolumn}
\newcommand{\kms}{km~s$^{-1}$}
\newcommand{\Te}{$T_{\rm e}$}

\newcommand{\sunn}{$_{\odot}$}
\newcommand{\HI}{H{\sc i}}
\DeclareRobustCommand{\ion}[2]{%
\relax\ifmmode
\ifx\testbx\f
{\mathrm{#1\,\textsc{#2}}}\else
{\mathrm{#1\,\mathsc{#2}}}\fi
\else\textup{#1\,{\mdseries\textsc{#2}}}%
\fi}

\title[New extreme LSBDs in the Lynx-Cancer void]
{Study of galaxies in the Lynx-Cancer void. -- III. New extreme LSB dwarf
galaxies
}
\author[S.A.~Pustilnik, J.-M.~Martin, A.L.~Tepliakova, A.Y.~Kniazev]
{S.A. Pustilnik,$^{1,5}$\thanks{E-mail: sap@sao.ru (SAP)}
J.-M. Martin,$^{2}$
A.L. Tepliakova,$^{1}$
A.Y. Kniazev$^{3,4}$ \\
\rule{-4pt}{20pt}
$^1$ Special Astrophysical Observatory of RAS, Nizhnij Arkhyz,
  Karachai-Circassia 369167, Russia\\
$^2$ GEPI and Station de radioastronomie, Observatoire de Paris, 5 place Jules Janssen, 92190 Meudon, France  \\
$^3$ South African Astronomical Observatory, PO Box 9, 7935 Observatory,
   Cape Town, South Africa\\
$^4$ Southern African Large Telescope Foundation, PO Box 9, 7935 Observatory,
   Cape Town, South Africa\\
$^5$ Isaac Newton Institute of Chile, SAO branch, Nizhnij Arkhyz, Russia}

\begin{document}

\label{firstpage}

\date{Accepted 2011 June 29. Received in original form  2011 April 28}

\pagerange{\pageref{firstpage}--\pageref{lastpage}} \pubyear{2011}

\maketitle

\begin{abstract}

We present the results of the complex study of the low surface brightness
dwarf (LSBD) gas-rich galaxies J0723+3621, J0737+4724 and J0852+1350,
which reside in the nearby Lynx-Cancer
void. Their ratios $M$(\HI)/$L_{\rm B}$,
according to \HI\ data obtained with the Nan\c{c}ay Radio Telescope (NRT),
are respectively $\sim$3.9, $\sim$2, $\sim$2.6. For the two latter galaxies,
we derived oxygen abundance corresponding to the value of
12+$\log$(O/H)$\lesssim$7.3, using spectra from the Russian 6m telescope
(BTA) and from the Sloan Digital Sky Survey (SDSS) database.
We found two additional blue LSB dwarfs, J0723+3622 and J0852+1351, which
appear to be physical companions of J0723+3621 and J0852+1350 situated
at the projected distances of $\sim$12--13~kpc. The companion relative
velocities, derived from the BTA spectra, are $\Delta~V$ = +89~\kms\  and
+30~\kms\ respectively. The geometry and the relative orientation of orbits
and spins in these pairs indicate, respectively, prograde and polar
encounters for J0723+3621 and J0852+1350.
The NRT \HI\ profiles of J0723+3621 and J0723+3622 indicate a sizable
gas flow in this system. The SDSS $u,g,r,i$ images of the five dwarfs are
used to derive the photometric parameters and the exponential or Sersic
disc model fits. For three of them, the $(u-g),(g-r),(r-i)$ colours of the
outer parts, being compared with the \mbox{PEGASE} evolutionary tracks,
evidence for the dominance of the old stellar populations with ages of
$T \sim$(8--10)$\pm$3~Gyr. For J0723+3622 and J0737+4724, the outer region
colours appear rather blue, implying the ages of the oldest visible stars of
T$\lesssim$1--3~Gyr. The new LSB galaxies complement the list of the known
most metal-poor and `unevolved' dwarfs in this void, including DDO~68, SDSS
J0812+4836, SDSS~J0926+3343 and SAO~0822+3545.
This unique concentration of 'unevolved' dwarf galaxies in a small cell of
the nearby Universe implies a physical relationship between the slow
galaxy evolution and the void-type  global environment. We also compare  the
baryonic content of these LSBDs with predictions of the most updated
cosmological simulations.
\end{abstract}

\begin{keywords}
galaxies: dwarf -- galaxies: ISM -- galaxies: abundances --
galaxies: photometry --  galaxies: evolution  -- cosmology: large-scale
structure of Universe
\end{keywords}

\section[]{INTRODUCTION}
\label{sec:intro}

Voids in the large-scale distribution of matter are defined observationally
as large regions devoid of luminous (L $>$ L$^{*}$) massive galaxies.
Dwarf galaxies in voids can have significantly different star formation (SF)
and chemical enrichment histories from those of galaxies in denser
environments
\citep[see, e.g.,][ and references therein]{Peebles01,
Gottlober03, Hoeft06, Arkhipova07, Hahn07, Hahn09}. To address the effect
of voids on evolution of the lowest mass dwarf galaxies, we compiled a
sample of $\sim$80  late-type galaxies in one of the nearest voids, the
Lynx-Cancer void \citep[Paper~I --][]{PaperI}.

Due to the void's relative proximity ($D_{\rm centre}$ $\sim$18~Mpc),
galaxies in this volume selected for the SDSS spectroscopy, have the absolute
magnitudes as low as $M_{\rm B} = -(12-13)$.
Galaxies residing in the Lynx-Cancer void (as well as those in other voids),
are mostly dwarfs ($\sim$95\%, Paper~I). There exists the
well known correlation between galaxy luminosities and their central surface
brightnesses (SB) \citep[e.g.,][ and references therein]{Cross02}.
The deeper one probes the void galaxy population, the higher fraction of LSB
galaxies one expects. However, the severe observational selection effects
prevents a good completeness sample of the low-luminosity LSB galaxies.

In the SDSS, a relatively high surface brightness  cut-off has been
used to select the spectroscopic targets. Namely, the completeness of the SDSS
galaxy samples with the measured redshifts falls below 50\% for objects
with the half-light $\mu_{50,r} \geq$
23.5~mag~arcsec$^{-2}$ \citep[e.g.,][]{Blanton05,Geha06}. This limit for the
purely exponential face-on discs, roughly corresponds to the observed value
of the central SB in $B$-band -- $\mu_{0,B}\geq$ 23.2~mag~arcsec$^{-2}$
(Paper~I). Hence, the resulting SDSS galaxy samples with known redshifts
are biased against the LSB galaxies in the nearby Metagalaxy. The apparent
brightening due to the inclination effect and the presence of extra light
related to the SF regions (or bulges), will somewhat diminish the loss of
galaxies with $\mu_{0,B,c,i}$ $<$ 24~mag~arcsec$^{-2}$. However, for galaxies
with $\mu_{0,B,c,i} \geq$ 24~mag~arcsec$^{-2}$, the SDSS selection criterion
should lead to the substantial loss of LSB dwarf galaxies in the void galaxy
samples. Here the indexes $c$ and $i$ correspond to the Galaxy foreground
extinction and the inclination corrected values. Some of these missing
galaxies might form the youngest local galaxy population \citep{Zackrisson05}.

The LSB galaxies are believed to evolve slower than the brighter counterparts
with similar masses. This occurs due to their lower surface mass density and
the larger stabilisations effect of their DM halos that suppress the internal
perturbations and the related gas collapse and SF process. Their properties
are also related to the higher specific angular momentum \citep{Dalcanton97}.
In turn, as cosmological simulations show, the primordial halos with the
higher specific angular momentum collapse late
and their baryonic gas can form LSBGs with substantial delays.
Furthermore, the same properties of LSBGs can eliminate the effect of
collisions and tidal perturbations, significantly reducing the induced SF
in comparison to the processes in the similar high-SB galaxies
\citep[e.g., ][]{Mihos97}. On the other hand, \citet{Neil98} suggested that
the blue colours of the LSB dwarfs can be understood as a result of starbursts
triggered by distant/weak tidal encounters. Similarly, \citet{Schombert01}
argued that the blue colours of LSBGs are related to the weak SF bursts,
caused by encounters have occurred during the last five Gyr.

In order to better disentangle various effects of galaxy interactions,
one needs to know how the isolated  galaxies evolve. The study of isolated
galaxies became rather popular especially in the last years, e.g.,
AMIGA project (\mbox{http://amiga.iaa.es/p/1-amiga-home.htm}),
by the creation of new samples of isolated galaxies - the Local Orphan
Galaxies sample \citep{Orphan}, 2MIG sample \citep{2MIG} and the
multiwavelength study of the sample of the most isolated galaxies in the more
distant voids \citep{VGS,Kreckel11}.
Also, since galaxies are complex systems, the separating of the most
extreme (`purified' from the confusing factors) galaxies, in which one finds
the simplest physical and dynamical conditions, helps to confront
the numerical simulations of physical processes in galaxies with the real
objects and to verify the model assumptions.

In this context, the very gas-rich and very metal-poor LSB galaxies residing
in voids, can appear to be the simplest objects for the understanding and
modelling
of their dynamics, star formation and evolution. The understanding of
various aspects of their lives can be the good baseline for the
analysis of more complicated cases and might give the keys for the insights
into  a more common aggregates and with a more complex
structure (in respect of the baryon component mix, spiral waves and bars,
etc). Is the very rarefied environment, characteristic of voids,
important for the LSBG evolution? Can we find any significant difference
in the evolutionary status between the LSB dwarfs residing in voids and
in groups and/or in the general field?

During the systematic study of the dwarf galaxies in the Lynx-Cancer void,
we have already discovered several unusual objects, including the almost
completed merger in the very metal-poor galaxy DDO~68
\citep{DDO68,IT07,DDO68_sdss}, the interacting pair of the
blue compact dwarf (BCD) HS~0822+3542 and the LSB dwarf SAO~0822+3545
\citep{HS0822,SAO0822,Chengalur06}, and the very LSB dwarf SDSS J0926+3343
\citep{J0926}.
In this paper we present the study of three more the Lynx-Cancer
void genuine LSB gas-rich dwarfs. The galaxies SDSS J073728.47+472432.8 and
SDSS J085233.75+135028.3 (hereafter J0737+4724 and J0852+1350)
had originally their redshifts in the SDSS database, while the galaxy
SDSS J072301.42+362117.1 (hereafter J0723+3621) does not.
The latter has been  identified as a Lynx-Cancer void dwarf with the
SAO RAS 6m telescope (BTA) during the snap-shot observations dedicated to
the search for LSBDs in this void.
We also present the radial velocity measurements (obtained with the
BTA) and the SDSS-based photometry of the two faint SDSS blue dwarf galaxies,
which are  located at the projected
distances of 2\arcmin\ and 2.6\arcmin\  from the studied void LSBDs.
For one of them we also discuss the NRT \HI\ data.

The paper is organised as follows. In Sec.~\ref{sec:obs} we
describe the observations, the SDSS data and the data reduction. Sec.
~\ref{sec:results} presents the results of observations and their analysis.
In Sec.~\ref{sec:dis} we discuss the results and their implications in the
broader context, comparing properties of the presented galaxies with
those of more common galaxy samples, and summarise our conclusions.

\section[]{OBSERVATIONS AND DATA REDUCTION}
\label{sec:obs}

\subsection{NRT \HI\ observations}
\label{NRT}

The \ion{H}{i}-observations with the
Nan\c {c}ay\footnote{
The Nan\c {c}ay Radioastronomy Station is part of the
Observatoire de Paris and is operated by the Minist\`ere de l'Education
Nationale and Institut des Sciences de l'Univers of the Centre National
de la Recherche Scientifique.}
radio telescope (NRT) with a collecting area of 200$\times$34.5~m$^2$ are
characterised by a half-power beam width (HPBW) of
3.7\arcmin~(East-West)$\times$22\arcmin~(North-South) at
declination $\delta$=0\degr\
(see also \verb|http/www.obspm-nancay.fr/en|).
The data were acquired during July 2009 - October  2010, with the total time
on-source of $\sim$4 hours for J0737+4724, $\sim$3 hours - for J0852+1350,
J0723+3621 and  J0723+3622. We used the antenna/receiver system F.O.R.T.
(Foyer Optimise pour le Radio Telescope) \citep[e.g.,][]{FORT} with the
improved overall sensitivity.
The system temperature was $\sim$35 K for both the horizontal and vertical
linear polarisations of a dual-polarisation receiver.
The gain of the telescope was 1.5 K~Jy$^{-1}$ at declination $\delta$=0\degr.
The 8192-channel correlator was used covering a total bandwidth of 12.5 MHz.
The total velocity range covered
was about 2700~\kms, with the channel spacing of 1.3~\kms\ before smoothing.
The observations consisted of separate cycles of `ON' and `OFF' integrations,
each of 40 seconds in duration. `OFF' integrations were acquired at the
target declination, with the East R.A. offset of
$\sim$15\arcmin~$\times$ cos($\delta$).
For more detail see the description in \citet{NRT_07}.

The data were reduced using the NRT standard programs NAPS and SIR, written
by the telescope staff (see description on
\mbox{http://www.nrt.obspm.fr}).
Horizontal and vertical polarisation spectra were calibrated and processed
independently and then averaged together. The error estimates were calculated
following to \citet{Schneider86}. The baselines were generally well-fit by a
third order or lower polynomial  and were subtracted out.

\subsection{6m telescope spectral observations and SDSS spectrum}
\label{BTA}

\begin{table*}
\begin{center}
\caption{Journal of the 6\,m telescope observations}
\label{Tab1}
\begin{tabular}{llrccccccc} \\ \hline \hline
\MC{1}{c}{ Name }       &
\MC{1}{c}{ Date }       &
\MC{1}{c}{ Expos. }   &
\MC{1}{c}{ Wavelength [\AA] } &
\MC{1}{c}{ Dispersion } &
\MC{1}{c}{ Spec.resol. } &
\MC{1}{c}{ Seeing }     &
\MC{1}{c}{ Airmass }     &
\MC{1}{c}{ Grism }       &
\MC{1}{c}{ Detector }    \\

\MC{1}{c}{ }       &
\MC{1}{c}{ }       &
\MC{1}{c}{ time [s] }    &
\MC{1}{c}{           } &
\MC{1}{c}{ [\AA/pixel] } &
\MC{1}{c}{ FWHM(\AA) } &
\MC{1}{c}{ [arcsec] }    &
\MC{1}{c}{          }    &
\MC{1}{c}{          }    &
\MC{1}{c}{          }     \\

\MC{1}{c}{ (1) } &
\MC{1}{c}{ (2) } &
\MC{1}{c}{ (3) } &
\MC{1}{c}{ (4) } &
\MC{1}{c}{ (5) } &
\MC{1}{c}{ (6) } &
\MC{1}{c}{ (7) } &
\MC{1}{c}{ (8) } &
\MC{1}{c}{ (9) } &
\MC{1}{c}{ (10) } \\
\hline
\\[-0.3cm]
J0723+3621  & 2009.12.23  & 1$\times$600 & $ 3700-6800$ & 2.1 & 12.0 & 2.0 & 1.13 & VPHG550G  & 2K$\times$2K  \\
J0723+3621  & 2010.11.09  & 2$\times$1200& $ 3500-7500$ & 2.1 & 12.0 & 2.0 & 1.15 & VPHG550G  & 2K$\times$2K  \\
J0737+4724  & 2009.01.21  & 5$\times$900 & $ 3700-6000$ & 0.9 & 5.5  & 1.2 & 1.12 & VPHG1200G & 2K$\times$4K  \\
J0852+1350  & 2010.11.11  & 2$\times$600 & $ 6000-7200$ & 0.9 & 5.5  & 3.0 & 1.41 & VPHG1200R & 2K$\times$2K  \\
\hline \hline \\[-0.2cm]
\end{tabular}
\end{center}
\end{table*}

\begin{figure}
 \centering
 \includegraphics[angle=-0,width=5.5cm,clip=]{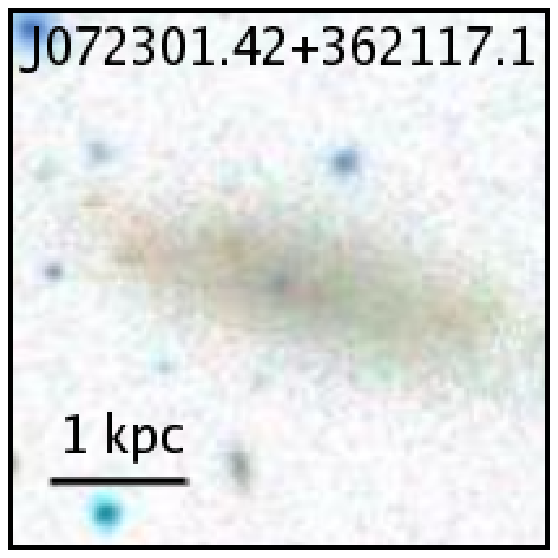}
 \includegraphics[angle=-0,width=5.5cm,clip=]{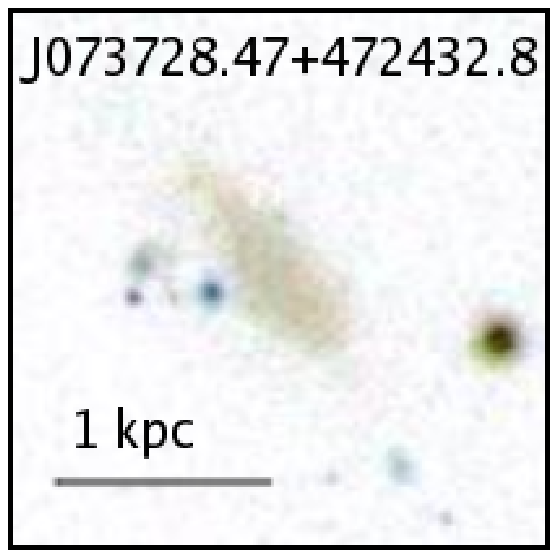}
 \includegraphics[angle=-0,width=5.5cm,clip=]{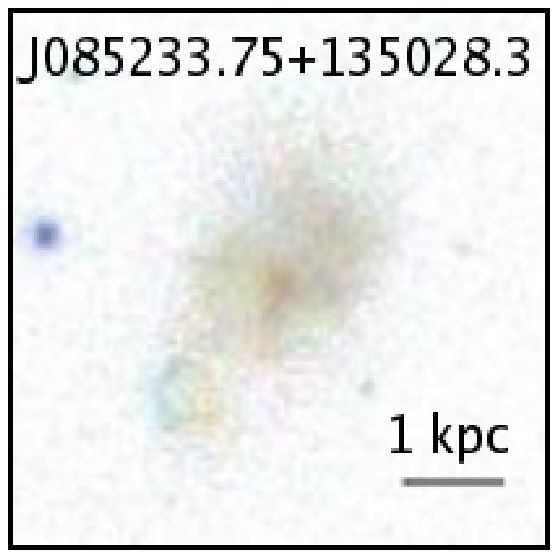}
  \caption{\label{fig:sdss_charts1}
  Finding charts of the three studied Lynx-Cancer LSBDs, obtained with the
SDSS Navigation Tool. N is up, E is to the left. The side measures
$\sim$50\arcsec. Horizontal bar in each figure corresponds to the linear
size of 1 kpc.
  {\bf Top panel:} SDSS J0723+3621.
  {\bf Middle panel}: SDSS J0737+4724.
{\bf Bottom panel}: SDSS J0852+1350. The `ring' structure on the SE edge
is reminiscent of the void galaxy DDO~68 \citep{DDO68}.
}
\end{figure}

The long-slit spectral observations of galaxies J0723+3621, J0737+4724 and
J0852+1350  (see their main parameters in Table~\ref{tab:param} and their
SDSS  finding charts -- in Fig.~\ref{fig:sdss_charts1})
were conducted with the multimode instrument SCORPIO \citep{SCORPIO}
installed at the prime focus of the SAO 6\,m telescope (BTA) on the
nights of 2009 January 21 and December 23; 2010 November 9 and 11.
The grisms VPHG550G, VPHG1200G and VPHG1200R  were used with either
the 2K$\times$2K CCD detector EEV~42-40 or 2K$\times$4K CCD detector
EEV~42-90. See details in Journal of observation in Table~\ref{Tab1}.
The scale along the slit (after binning) was 0\farcs36 pixel$^{-1}$
in all cases.
The object spectra were complemented before or after by the reference
spectra of He--Ne--Ar lamp for the wavelength calibration. The spectral
standard star Feige~34 \citep{Bohlin96} was observed during the nights for
the flux calibration.

\begin{figure}
 \centering
 \includegraphics[angle=-0,width=7.5cm,clip=]{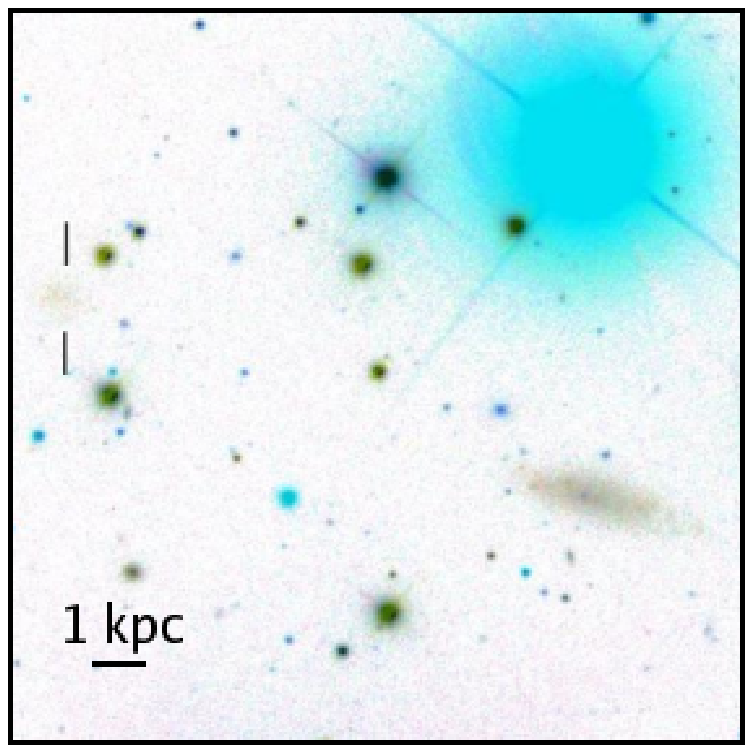}
 \includegraphics[angle=-0,width=7.5cm,clip=]{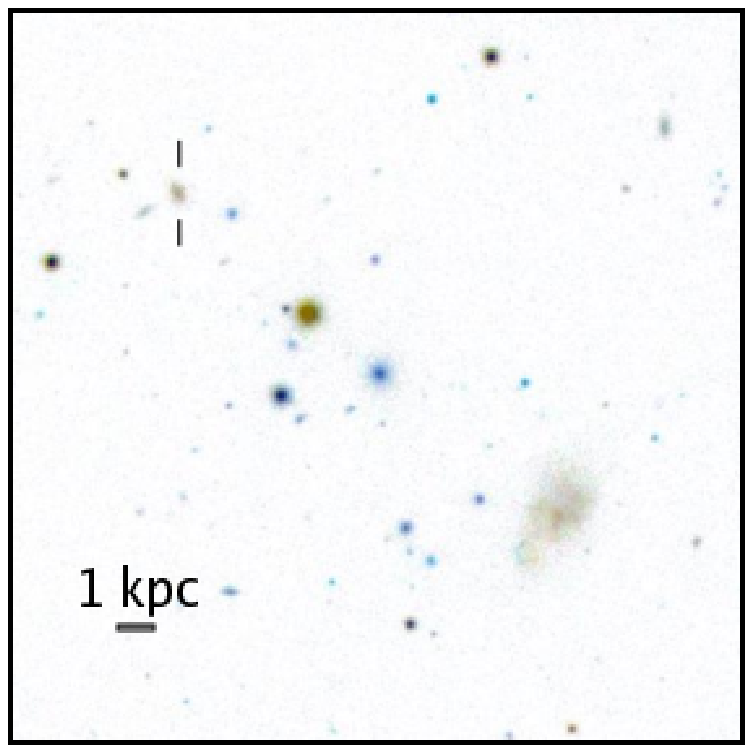}
  \caption{\label{fig:sdss_charts2}
  Finding charts of blue dwarf companions (marked by vertical bars)
of the Lynx-Cancer LSBDs (which are also seen in SW corners), obtained with
the SDSS Navigation Tool.
  N is up, E is to the left. Side measures $\sim$200\arcsec.
  Horizontal bar corresponds to linear size of 1 kpc.
  {\bf Top panel:} The void LSBD SDSS J0723+3621 along with its companion
SDSS J0723+3622.
{\bf Bottom panel}: The void LSBD SDSS J0852+1350 along with its companion
SDSS J0852+1351.
}
\end{figure}

For J0723+3621, the long slit was positioned along the major axis at the
position angle PA=69\degr.  For the second observation it was displaced from
the detector centre so that the neighbour galaxy J0723+3622 was also on
the slit.
For J0737+4724, the slit was positioned on two H{\sc ii} regions ("a"
and  "b") at the south-west edge of the galaxy body (PA=--111\degr).  These
H{\sc ii} regions (with a distance of $\sim$4\arcsec\ in between) were
identified through the one-minute exposure acquisition images of this galaxy
with a SED665 filter (centre at $\lambda$6622~\AA, FWHM=191~\AA).
For J0852+1350, the slit was positioned over the brightest
`central' knot in this galaxy (same as for the SDSS spectrum) and to include
the brightest part of the candidate companion galaxy J0852+1351 at
$\sim$2\arcmin\ NE (PA=44\degr).  We also used the SDSS spectrum of
J0852+1350 (spID=53815-2430-597) for analysis. It has the useful wavelength
range 3800-9000~\AA\ with the typical FWHM of emission lines of $\sim$3\AA.
The
spectrum was acquired in 3\arcsec\ diaphragm, positioned on the brightest
knot near the galaxy centre (see Fig.~\ref{fig:sdss_charts1}).

All spectral data reduction and emission line
measurements were performed similar to that described in \citet{DDO68}.
Namely, the standard pipeline with the use of IRAF\footnote{IRAF: the Image
Reduction and Analysis Facility is distributed by the National Optical
Astronomy Observatory, which is operated by the Association of Universities
for Research in Astronomy, Inc. (AURA) under cooperative agreement with the
National Science Foundation (NSF).}
and {\tt MIDAS}\footnote{MIDAS is an acronym for the European Southern
Observatory package -- Munich Image Data Analysis System. }
was applied for the reduction of long-slit spectra, which included the
following steps: removal of cosmic ray hits,
bias subtraction,  flat-field correction, wavelength
calibration, night-sky background subtraction. Then, using the data on the
spectrophotometry standard star, all spectra were transformed to absolute
fluxes. The emission line intensities with their errors were measured in
the way described in detail in \citet{SHOC}.

\subsection{Imaging data from the SDSS database}

The SDSS \citep{York2000} is well suited for
photometric studies of various galaxy samples due to its homogeneity, area
coverage, and depth (SDSS Project Book\footnote{
http://www.astro.princeton.edu/PBOOK/welcome.htm}).
SDSS is an imaging and spectroscopic survey that covers about
one-quarter of the Celestial Sphere. The imaging data are collected in drift
scan mode in five bandpasses \citep[$u, \ g, \ r, \ i$, and $z$;][]{SDSS_phot}
using mosaic CCD camera \citep{Gunn98}. An automated image-processing system
detects astronomical sources and measures their photometric and astrometric
properties \citep{Lupton01,SDSS_phot1,Pier03} and identifies candidates for
multi-fibre spectroscopy.
At the same time, the pipeline-reduced SDSS data can be used, if
necessary, in order to get independent photometry \citep[e.g.,][]{Kniazev04}.
For our current study the images in the respective filters were retrieved from
the SDSS Data Release 7 \citep[DR7;][]{DR7}.

Since the SDSS provides users with the fully reduced images, the only
additional step we needed to perform (apart from the photometry in round
diaphragms) was the background subtraction. For this, all bright stars were
removed from the images. After that the studied object was masked and the
background level within this mask was approximated with the package {\tt aip}
from {\tt MIDAS}.  The method and the related programs are
described in more detail in \cite{Kniazev04}.
To transform instrumental fluxes in diaphragms to stellar magnitudes, we
used the photometric system coefficients defined in SDSS for the used fields.
The accuracy of zero-point determination was  $\sim$0.01 mag in all
filters.

\section[]{RESULTS}
\label{sec:results}

\subsection[]{Spectra and oxygen abundance}

\subsubsection{J0723+3621 and J0723+3622}

There are no spectra for J0723+3621 in the SDSS DR7 database. We obtained
a snap-shot (with the acquisition time of 10 minutes) spectrum at BTA in
the frame of the Lynx-Cancer void new LSBD search program.
The spectrum was acquired along the galaxy major axis. It covers two compact
faint peripheral H{\sc ii} regions  and a brighter extended one close to the
geometric centre of the galaxy, but substantially displaced from the
brightest region of the stellar continuum. The S-to-N ratio in the extracted
1D spectrum of this the brightest H{\sc ii} region is low.

The second spectrum (with the total acquisition time of 40 minutes) was
acquired also along the galaxy major axis, at PA=69\degr, but the galaxy was
displaced from the centre of CCD detector by $\sim$1.3\arcmin, in order
to place the nearby galaxy J0723+3622 on the slit on the usable region of
the CCD detector. This allowed the relative velocity of the ionised gas
in the both galaxies to be measured. See discussion in \ref{sec:companions}.
Despite a factor of $\sim$2 higher S-to-N ratio for this spectrum of
J0723+3621, the resulting accuracy of the derived physical parameters and
the O/H value is still low. We do not discuss this further, but the main
strong oxygen line ratios are consistent with the value of
12+$\log$(O/H)$\lesssim$7.5.

\begin{figure*}
 \centering
 \includegraphics[angle=-90,width=6.8cm]{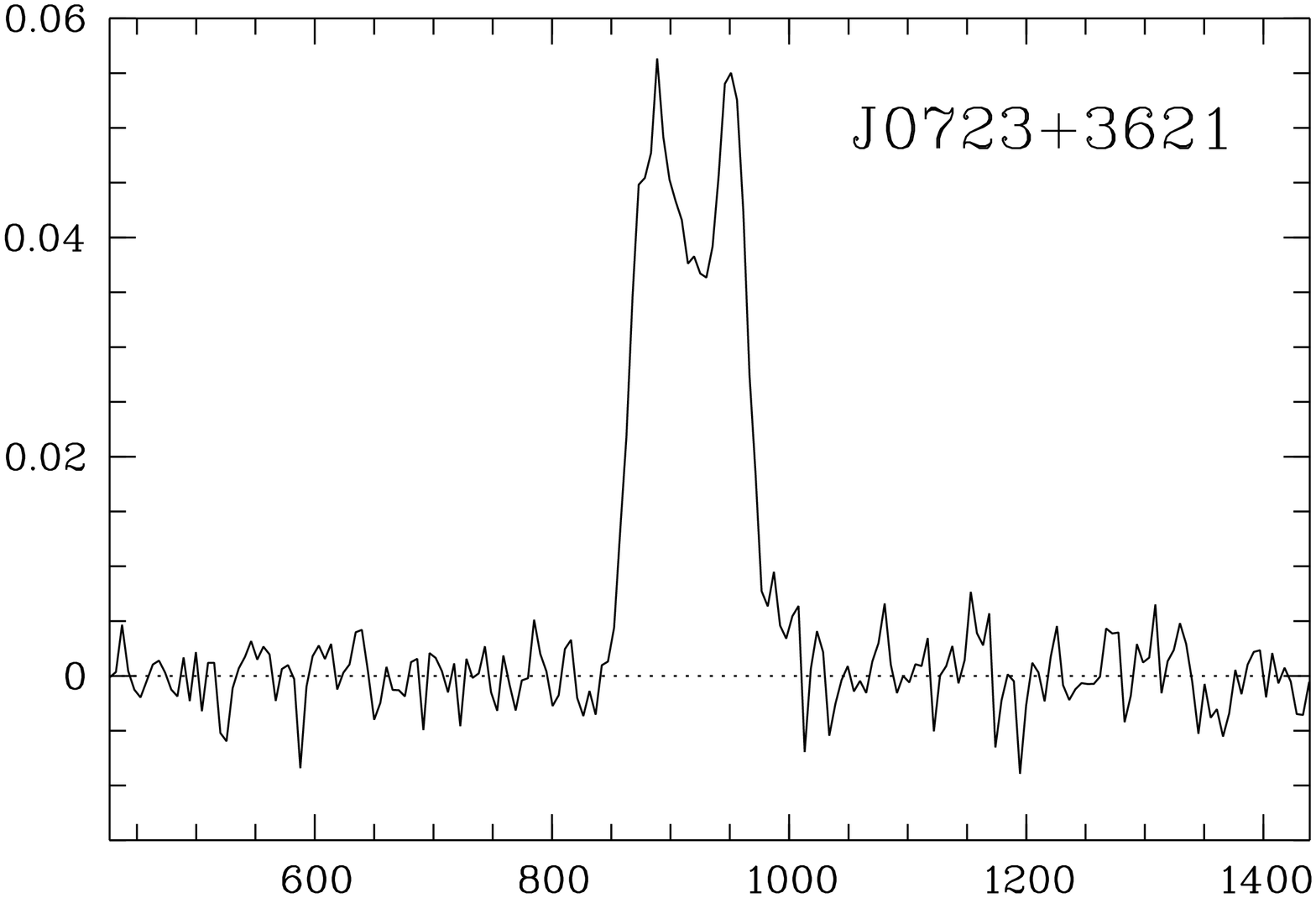}
 \includegraphics[angle=-90,width=6.8cm]{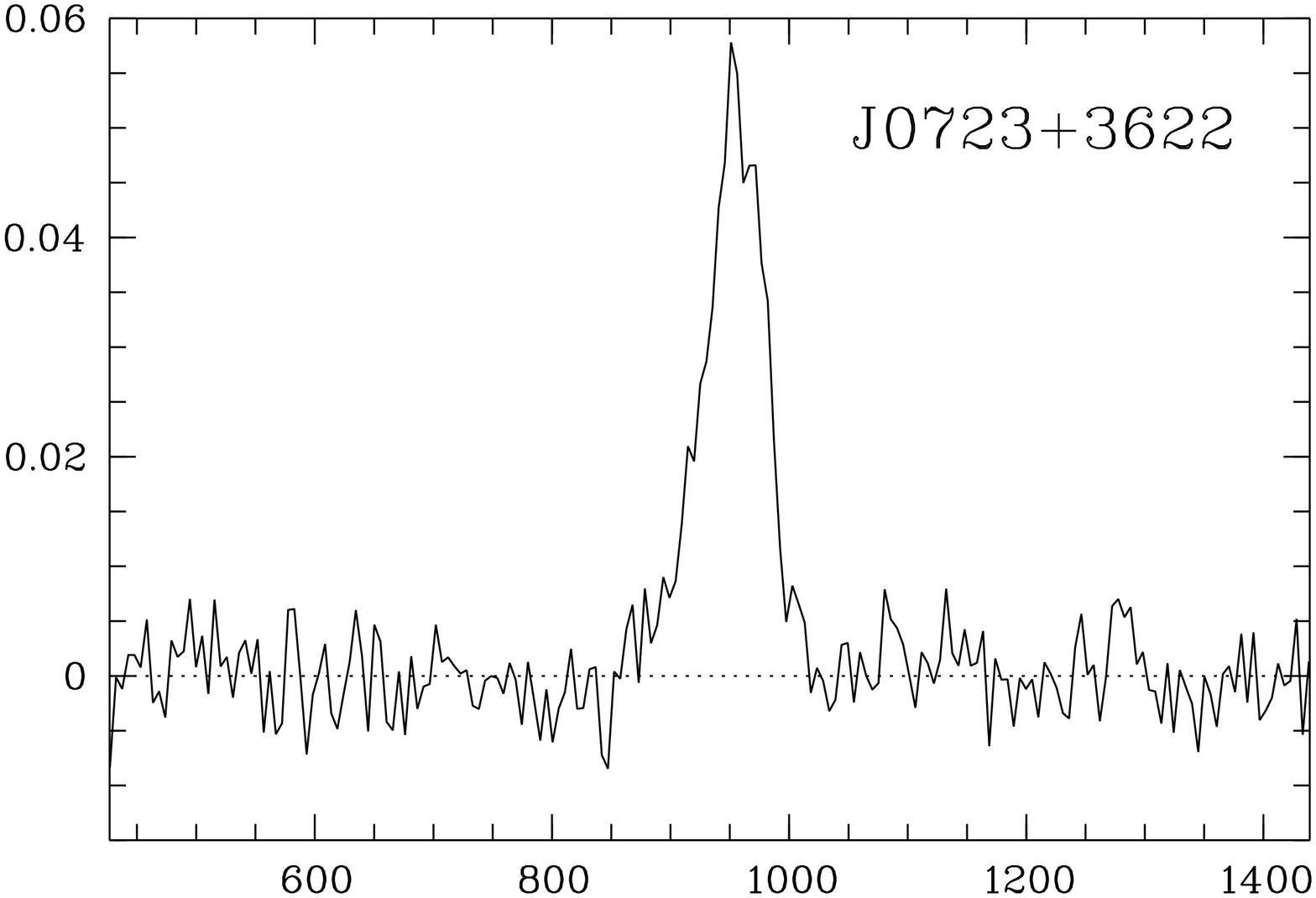}
 \includegraphics[angle=-90,width=6.8cm]{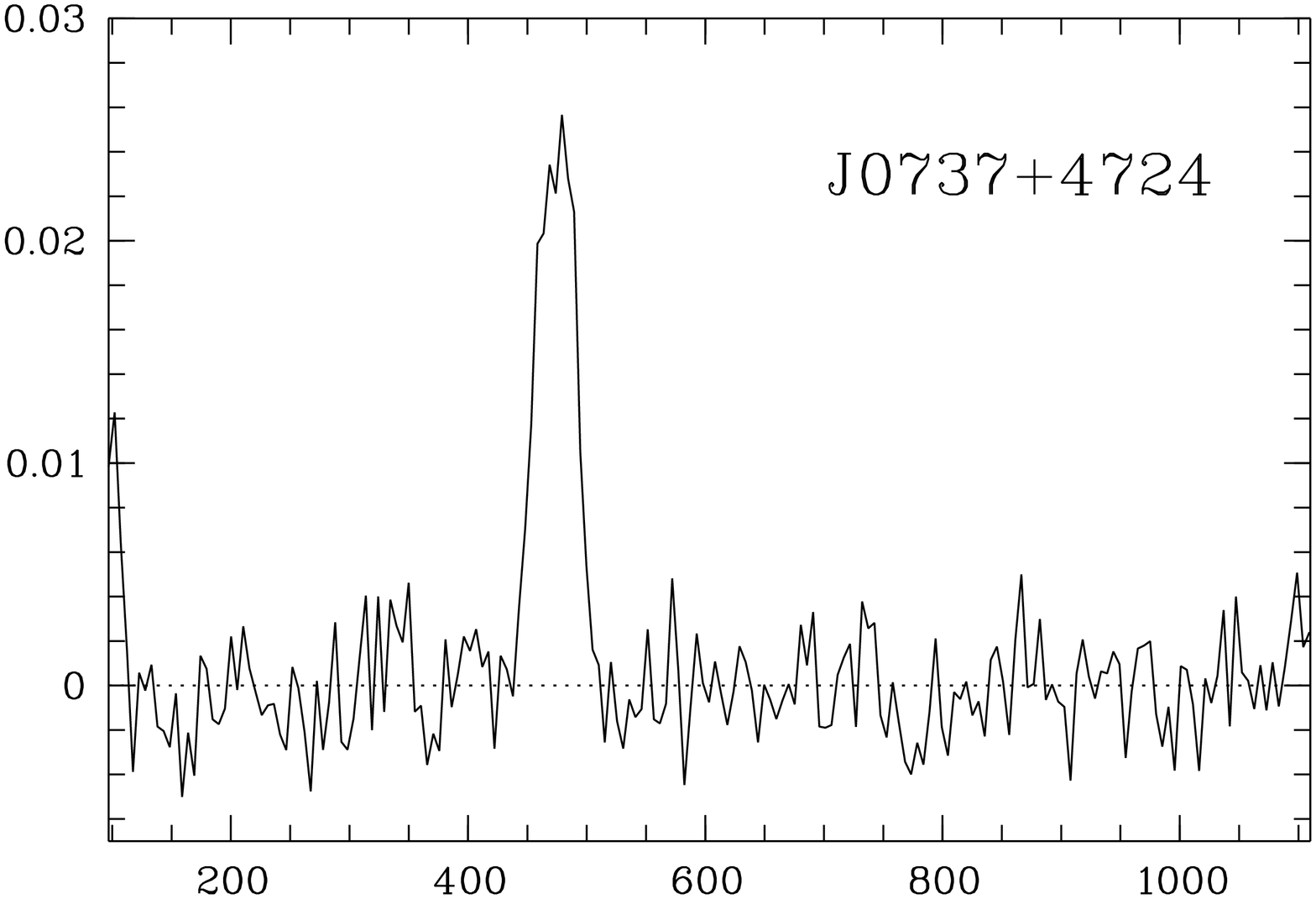}
 \includegraphics[angle=-90,width=6.8cm]{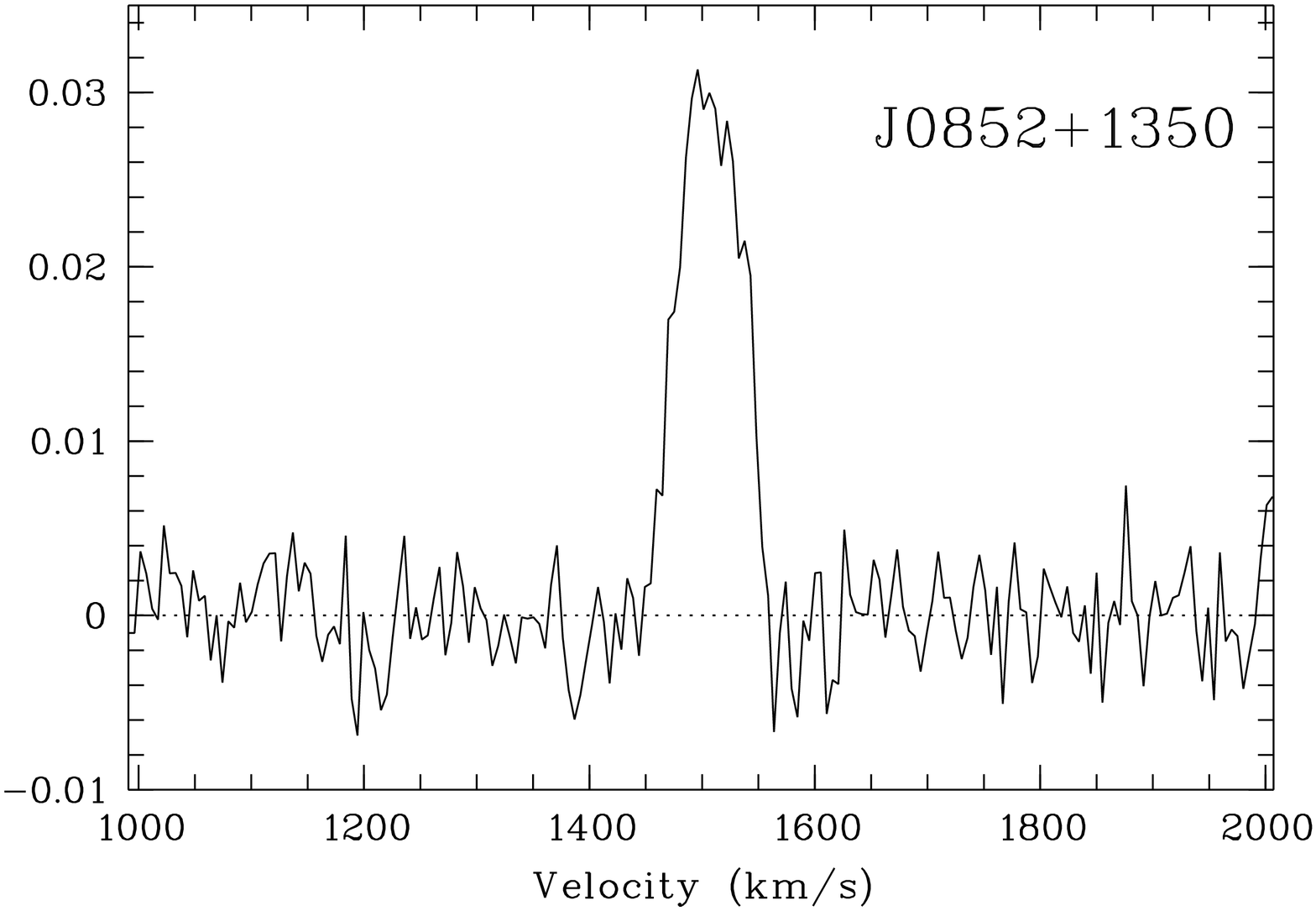}
  \caption{\label{fig:HI} H{\sc i} profiles of the studied Lynx-Cancer LSBDs
obtained with NRT. X-axis presents the heliocentric velocity in \kms.
Y-axis shows the galaxy H{\sc i} emission flux density in Jy.
{\bf Top left:}  Profile for NRT pointing at J0723+3621.
{\bf Top right:} Profile for NRT pointing at J0723+3622.
{\bf Bottom left:} J0737+4724.
{\bf Bottom right:} J0852+1350.
}
\end{figure*}

\subsubsection{J0737+4724}

\begin{table}
\centering{
\caption{Line intensities in spectra of J0737+4724 and J0852+1350}
\label{tab:Intens1}
\begin{tabular}{lcccc} \hline \hline
\rule{0pt}{10pt}
& \MC{2}{c}{J0737+4724} & \MC{2}{c}{J0852+1350}  \\ \cline{2-3} \cline{4-5}
\rule{0pt}{10pt}
$\lambda_{0}$(\AA) Ion                    &
$F$/$F$(H$\beta$)&$I$/$I$(H$\beta$) &
$F$/$F$(H$\beta$)&$I$/$I$(H$\beta$) \\ \hline

3727\ [O\ {\sc ii}]\            &  99$\pm$65 & 95$\pm$73  & (60$\pm$24)& (79$\pm$32)\\
4101\ H$\delta$\                &  17$\pm$5  & 29$\pm$10  &  22$\pm$4   &  26$\pm$7   \\
4340\ H$\gamma$\                &  34$\pm$8  & 48$\pm$15  &  50$\pm$5   &  56$\pm$7   \\
4363\ [O\ {\sc iii}]\           &      ...   &     ...    &  9$\pm$3    &  10$\pm$3   \\
4861\ H$\beta$\                 & 100$\pm$12 & 100$\pm$15 &  100$\pm$6  &  100$\pm$7  \\
4959\ [O\ {\sc iii}]\           & 59$\pm$10  & 51$\pm$10  &  82$\pm$5   &  80$\pm$5  \\
5007\ [O\ {\sc iii}]\           & 177$\pm$18 & 151$\pm$18 &  263$\pm$13 &  254$\pm$13  \\
5876\ He\ {\sc i}\              & 20$\pm$11  & 16$\pm$10  &  7$\pm$3    &  6$\pm$2   \\
6548\ [N\ {\sc ii}]\            &      ...   &     ...    &  1$\pm$1    &  1$\pm$1   \\
6563\ H$\alpha$\                &      ...   &     ...    &  365$\pm$18 &  272$\pm$14 \\
6584\ [N\ {\sc ii}]\            &      ...   &     ...    &  4$\pm$1    &   3$\pm$1   \\
6716\ [S\ {\sc ii}]\            &      ...   &     ...    &  33$\pm$4   &  24$\pm$3   \\
6730\ [S\ {\sc ii}]\            &      ...   &     ...    &  19$\pm$2   &  14$\pm$2   \\
7320\ [O\ {\sc ii}]\            &      ...   &     ...    &  ...        &   ...           \\
7330\ [O\ {\sc ii}]\            &      ...   &     ...    &  4$\pm$2    &   3$\pm$1    \\
& &  &  & \\
C(H$\beta$)\ dex                & \MC {2}{c}{0.14$\pm$0.16} & \MC {2}{c}{0.38$\pm$0.06}   \\
EW(abs)\ \AA\                   & \MC {2}{c}{4.50$\pm$0.68} & \MC {2}{c}{0.05$\pm$0.98}   \\
$F$(H$\beta$)$^a$\              & \MC {2}{c}{1.81$\pm$0.16} & \MC {2}{c}{8.81$\pm$0.79}   \\
EW(H$\beta$)\ \AA\              & \MC {2}{c}{ 28.1$\pm$2.5} & \MC {2}{c}{ 27.5$\pm$1.2}   \\
V$_{hel}$\ \kms\                & \MC {2}{c}{ 459$\pm$21}   & \MC {2}{c}{1586$\pm$30}     \\ \hline \hline
\MC{5}{l}{$^a$ in units of 10$^{-16}$ ergs\ s$^{-1}$cm$^{-2}$.}
\end{tabular}
 }
\end{table}

\begin{table}
\centering{
\caption{Derived Oxygen abundances }
\label{t:Chem1}
\begin{tabular}{lcc} \hline \hline
\rule{0pt}{10pt}
\rule{0pt}{10pt}
Value                                 & J0737+4724       & J0852+1350       \\
				      & semi-empir       & (T$_{\rm e,c}$)  \\  \hline
$T_{\rm e}$(OIII)(10$^{3}$~K)\        & 19.79$\pm$2.24   & 22.20$\pm$5.13   \\
$T_{\rm e}$(OII)(10$^{3}$~K)\         & 14.94$\pm$2.25   & 16.24$\pm$5.25   \\
$N_{\rm e}$(SII)(cm$^{-3}$)\          &   10$\pm$10~~    &  10$\pm$10~~     \\
&   & \\
O$^{+}$/H$^{+}$($\times$10$^{-5}$)\   & 0.872$\pm$0.782  & 0.636$\pm$0.539  \\
O$^{++}$/H$^{+}$($\times$10$^{-5}$)\  & 0.883$\pm$0.226  & 1.166$\pm$0.520  \\
O/H($\times$10$^{-5}$)\               & 1.755$\pm$0.814  & 1.802$\pm$0.749  \\
12+log(O/H)\                          & ~7.24$\pm$0.20~  & ~7.26$\pm$0.18~  \\
\hline \hline
\end{tabular}
 }
\end{table}

\begin{figure}
  \centering
 \includegraphics[angle=-90,width=7.5cm, clip=]{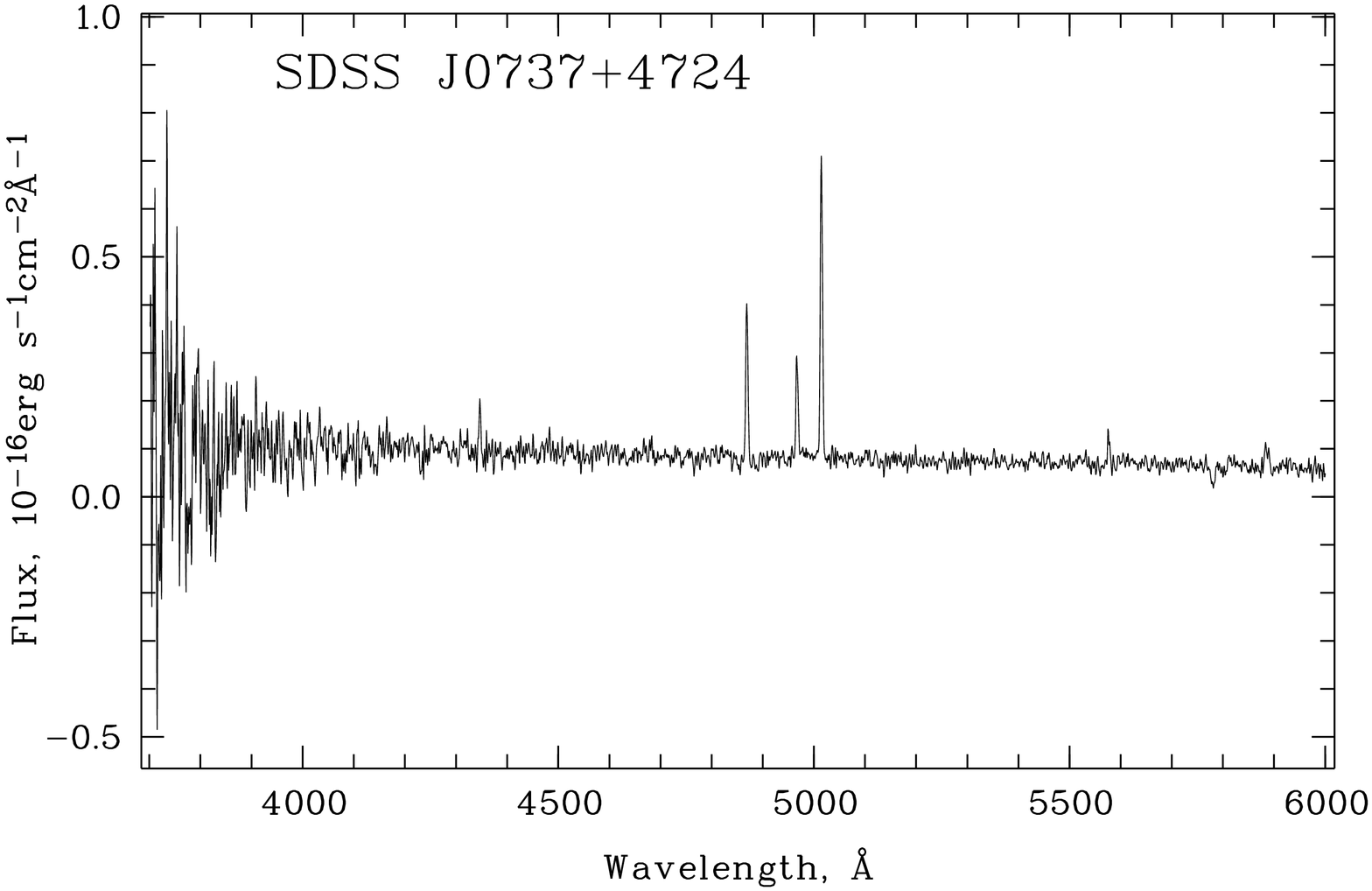}
 \includegraphics[angle=-90,width=7.5cm, clip=]{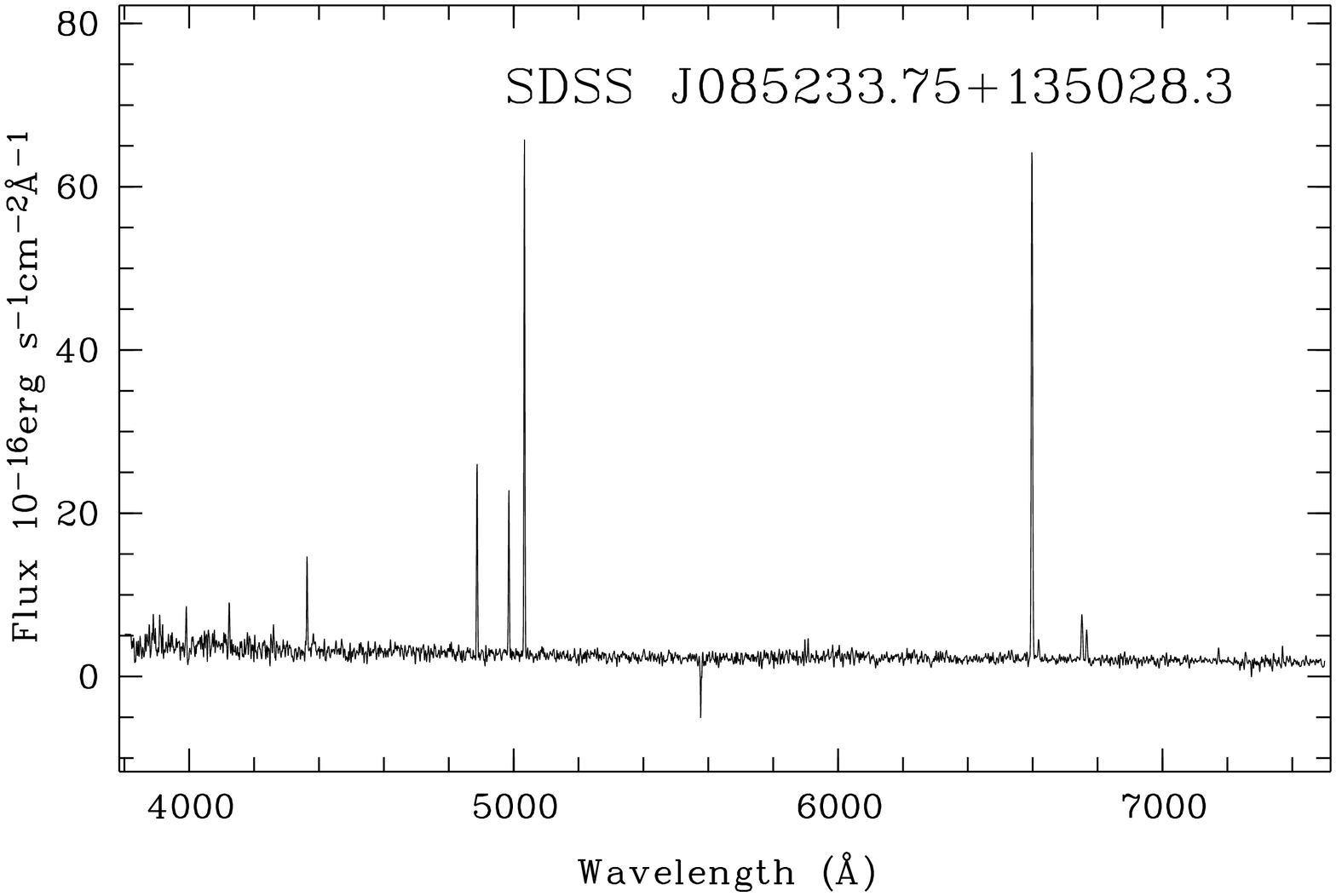}
  \caption{
Spectra of two Lynx-Cancer void LSBDs.
{\bf Top panel:} The BTA spectrum of J0737+4724 in the range
3700--6000~\AA\ with resolution of 5.5~\AA.
{\bf Bottom panel:} The SDSS spectrum of J0852+1350 in the range
3800--7500~\AA\  with resolution of $\sim$3~\AA.
}
	\label{fig:spectra}
 \end{figure}

The BTA spectrum of region {\bf a} is shown in Fig.~\ref{fig:spectra}.
Due to the strong noise in the UV range, the line
[\ion{O}{ii}] $\lambda$3727 has rather low S-to-N ratio.
In Table~\ref{tab:Intens1}, we present the relative (in respect of
F(H$\beta$)),  line intensities
F($\lambda$) of all relevant emission lines measured in the spectrum,
integrated over the region of 7.5\arcsec\ along the slit,
and I($\lambda$), corrected for the foreground extinction C(H$\beta$) and
the equivalent widths of the underlying Balmer absorption lines EW(abs).
Despite the rather long integration time, the S-to-N ratio in the presented
spectrum is quite low because the emission-line region is very faint. The
principal faint line [\ion{O}{iii}] $\lambda$4363 was not detected. The
intensity line ratios for [\ion{O}{iii}] $\lambda\lambda$4959,5007 and
H$\beta$ are indicative of low metallicity. In the case when the
[\ion{O}{iii}] $\lambda$4363 is barely seen or undetected, in the
low-metallicity regime, \citet{IT07} proposed the so-called semi-empirical
method of O/H determination, which employs the relation between the sum of
relative intensities of lines
[\ion{O}{iii}] $\lambda\lambda$4959,5007 and [\ion{O}{ii}] $\lambda$3727 and
\Te\ \citep[see the similar method for the high-metallicity regime in][]
{Pagel79,Shaver83}.  This method was tested on several of the most metal-poor
\ion{H}{ii}-regions in \citet{IT07} and also by us on our own data.
The O/H values derived by this method, appeared to be consistent, within
rather small errors (0.07~dex), with those derived via the direct
\Te-method.
The calculations of O/H were performed on the same formulae as described
above in the direct method. But instead of the \Te\ derived through the
intensity ratio of [\ion{O}{iii}] lines $\lambda$4363 and
$\lambda\lambda$4959,5007,  we adopt the \Te\ derived with the empirical
formula from \citet{IT07}.
In Table~\ref{t:Chem1} we present  the electron temperatures for
zones of emission of O$^{++}$ and  O$^{+}$, the adopted electron
densities $N_{\rm e}$, and the ionic abundances of oxygen, along with the
total abundances derived for the above measured line intensities,
according to the scheme described in \citet{Kniazev08}.

\subsubsection{J0852+1350 and J0852+1351}
\label{obsJ0852}

In Table~\ref{tab:Intens1}, we present the line intensities F($\lambda$) of
all relevant emission lines measured in the SDSS spectrum of J0852+1350
(spID=2430-53815-597). The line [\ion{O}{ii}]$\lambda$3727 is out of the
range for the SDSS spectra with redshifts of $z <$0.025. As was suggested in
\citet{Kniazev03,SHOC}, we employ the variant of the classic \Te\ method in
which the line intensities of [\ion{O}{ii}]$\lambda\lambda$7320,7330 are
used.
Despite the principal line [\ion{O}{iii}]$\lambda$4363 is well seen, its
form and the sizable shift from the correct wavelength position indicate
that its intensity is affected by the noise.
Indeed, the formal estimate of the r.m.s. error of this line intensity after
the MIDAS deblending procedure (with H$\gamma$ line), gives the
uncertainty of $\sim$30\%. Furthermore, the derived by the classic
\Te-method the related temperatures appear to be non-physical (too high).
To proceed further,
we reduced the intensity of [\ion{O}{iii}]$\lambda$4363 by $\sim$30\%, that
corresponds to 1~$\sigma_{\rm noise}$. Therefore, the values of $T_{\rm e}$
in Table~\ref{t:Chem1} (2-nd column) have relatively large errors.

The BTA spectrum of this galaxy in the range of 6000--7200~\AA\ was obtained
with the slit positioned on the brightest central knot and including the blue
faint galaxy at $\sim$2\arcmin\ NE. For the larger galaxy, no new information
was obtained on the relative line intensities in respect of that derived from
the SDSS spectrum. The moderately strong emission H$\alpha$ line
($EW \sim$38~\AA) was detected in the companion.
The measurements of the H$\alpha$-line wavelengths in the
both objects resulted in a  determination of their radial velocities:
$V_{\rm hel}$(J0852+1350)=1534$\pm$4~\kms; and $V_{\rm hel}$(J0852+1351)
=1565$\pm$22~\kms. The relative velocity of the fainter component
is +31$\pm$22~\kms.

\subsection[]{H{\sc i} parameters}

The profiles of the 21-cm \HI\ line emission in J0723+3621, J0723+3622,
J0737+4724 and J0852+1350, obtained with NRT, are shown in Fig.~\ref{fig:HI}.
Their observed and  derived parameters are summarised in
Table~\ref{tab:param}. The signal from the galaxies J0723+3621 and J0723+3622,
due to their proximity, should be affected by their mutual confusion.
Since the angular distance between the two galaxies is only 145.5\arcsec\
in R.A. and 0.93\arcmin\ in Declination, the signal of one galaxy
for the beam pointing to its neighbour, due to the beam off-set, is expected
to be $\sim$0.3 of its nominal value, if the H{\sc i} emission of each galaxy
is concentrated near its optical counterpart and has a typical size of
$\lesssim$1\arcmin. Having the independent information on the optical
velocities of the both components and the known (Gaussian) function of
the signal decrease due to the beam off-set, one can disentangle at first
approximation the contribution from the both galaxies.

The analysis of these \HI\ profiles taking into account for the mentioned
above \HI-flux decrease and for all a priori information from the optical
spectra, leads to the following model. The two velocity peaks correspond
to two spatially different \HI\ components. The first one, at
$V_{\rm hel}$=888~\kms\  is related to the larger, western LSBD J0723+3621.
But its centre is probably shifted westward by $\sim$23\arcsec (1.7~kpc).
The second component, with $V_{\rm hel}$=954~\kms, has approximately the same
\HI-flux as the first component. But its position is quite far from the
eastern component J0723+3622, since its apparent flux does not change
significantly for two pointings. The simplest explanation is that
the \HI\ component is situated somewhere in the middle
between the two LSBDs, that is it represents a gas flow or a tidal bridge.
The \HI\ integrated flux presented in Table~\ref{tab:param} for J0723+3621
gives the total flux of the system after the respective corrections for the
beam off-sets, assuming that due to the large difference in the component
luminosities, the \HI-mass of J0723+3622 contributes a small unknown fraction,
which probably does not exceed 10--15 per cent. Due to the large
uncertainties, we do not present the \HI\ parameters of J0723+3622, except
the velocity of the second component. To obtain a more detailed picture of
\HI\ in this system, we need a map with the
angular resolution of $\sim$10\arcsec\ or better.

To estimate the galaxies' global parameters, we adopted $D$=15.6, 10.4 and
23.1 Mpc for their distances, respectively  and the linear scales of 73,
52 and 112 pc in 1 arcsec.
The distances are derived from their $V_{\rm LG}$=885,
521 and 1360~\kms, the accepted Hubble constant
of 73~\kms~Mpc$^{-1}$ and the correction for the large negative peculiar
velocity discussed by \citet{Tully08}, which we adopt for these galaxies
as 264~\kms, 235~\kms\  and 322~\kms, respectively. See also Paper~I for
a more detailed discussion of the distance issue for the Lynx-Cancer void
galaxies. The \HI\ mass of each galaxy is determined by the well-known
formula for an optically thin \HI-line emission \citep{Roberts69}.

\subsection{Photometric properties and the age estimates}
\label{sect:age}

Based on the results of photometry and analysis of the SDSS images, we present
in Table~\ref{tab:photo} (the upper part) several model-independent
parameters such as the total magnitudes $g_{\rm tot}$, the related four
colour
indexes - $u-g$, $g-r$, $r-i$ and $i-z$, and the recalculated $B_{\rm tot}$.
In the other lines we give the {\it effective} angular radii at the SB =
25~mag~arcsec$^{-2}$ in $g$ and $r$ filters, the axial ratio $b/a$ at this
SB level (adopted from NED/SDSS), and the values of the Holmberg radii (at
SB = 26.5~mag~arcsec$^{-2}$) in $g$ and $r$ filters.

\begin{figure}
  \centering
 \includegraphics[angle=-90,width=8.0cm, clip=]{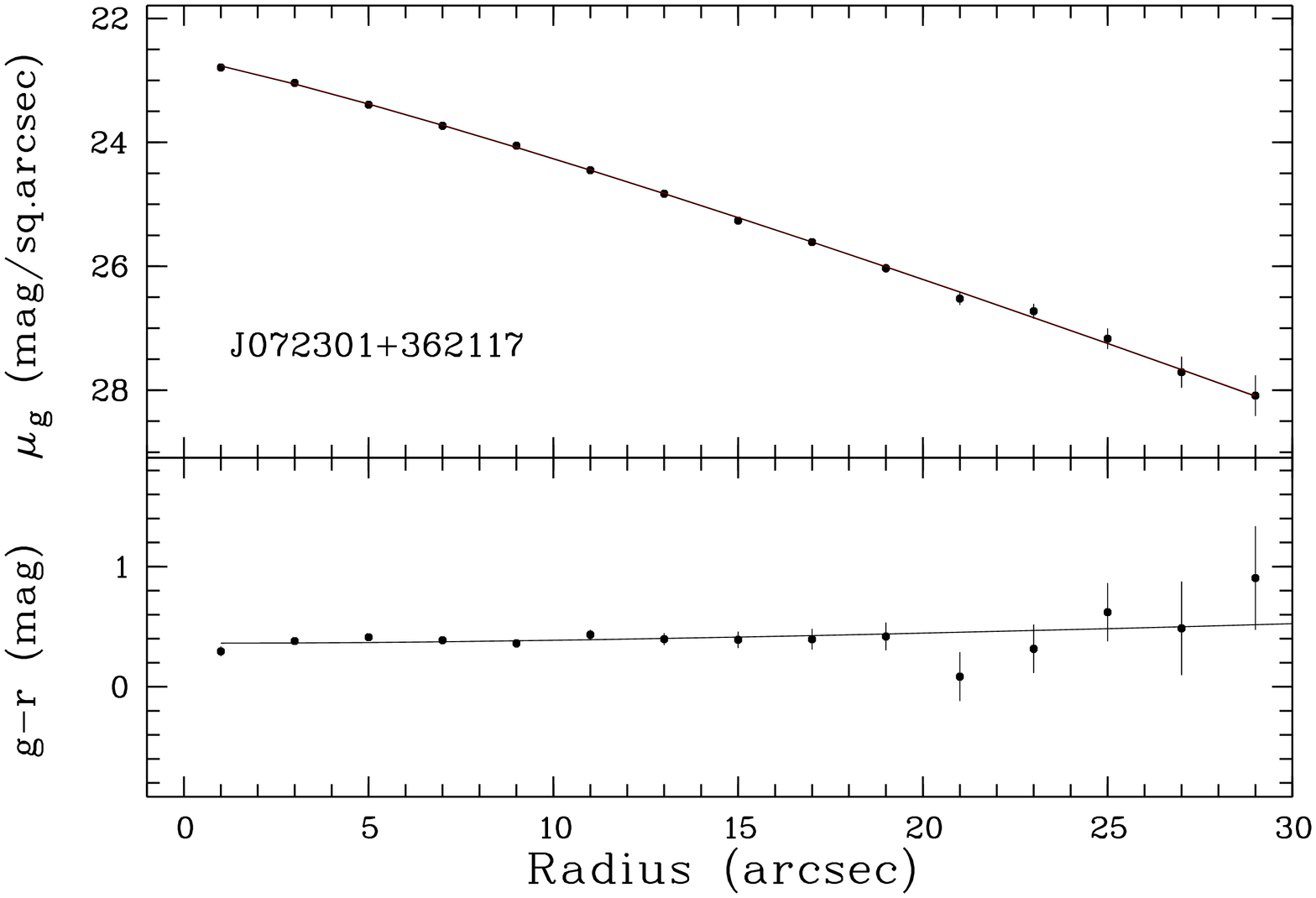}
 \includegraphics[angle=-90,width=8.0cm, clip=]{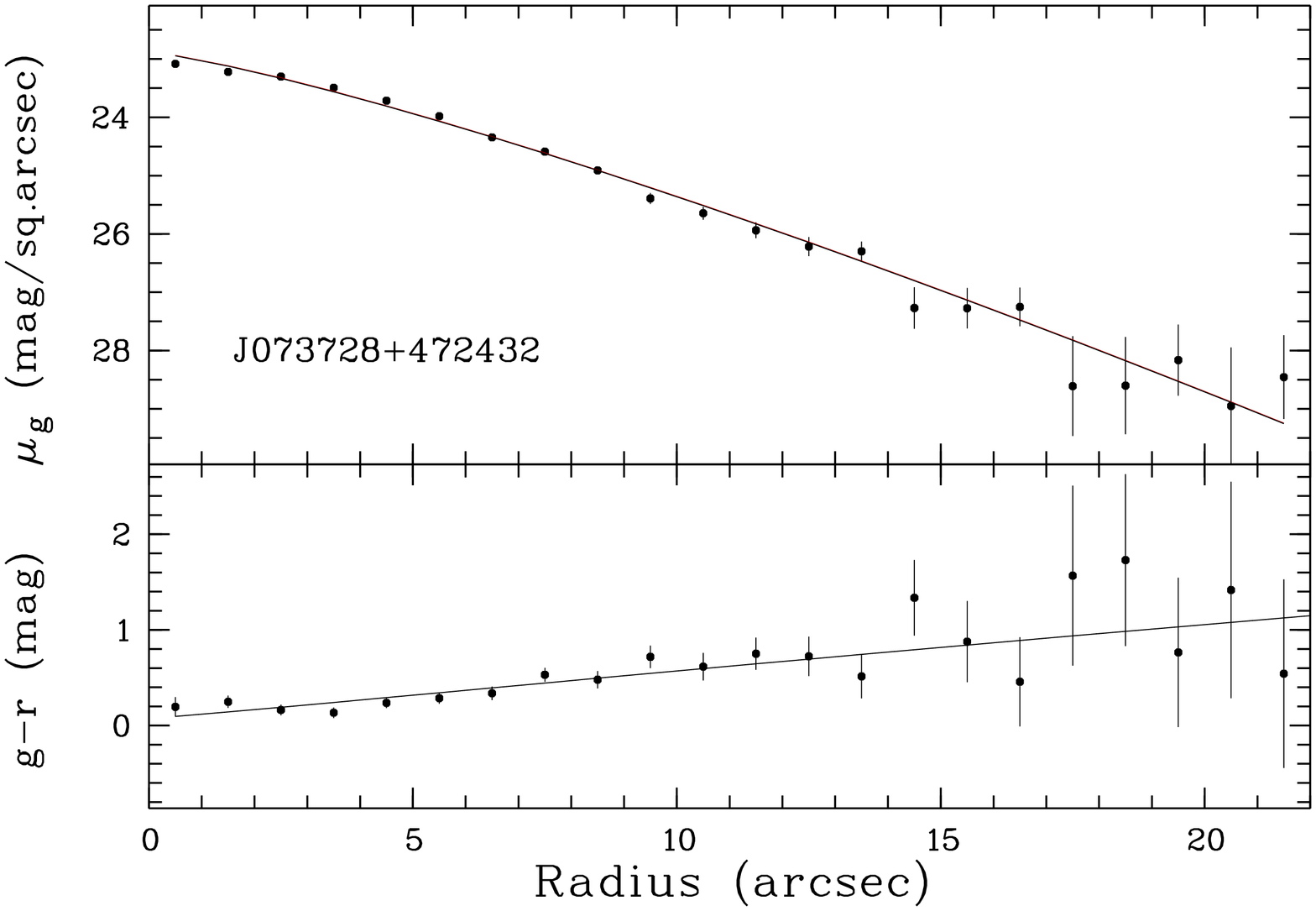}
 \includegraphics[angle=-90,width=8.0cm, clip=]{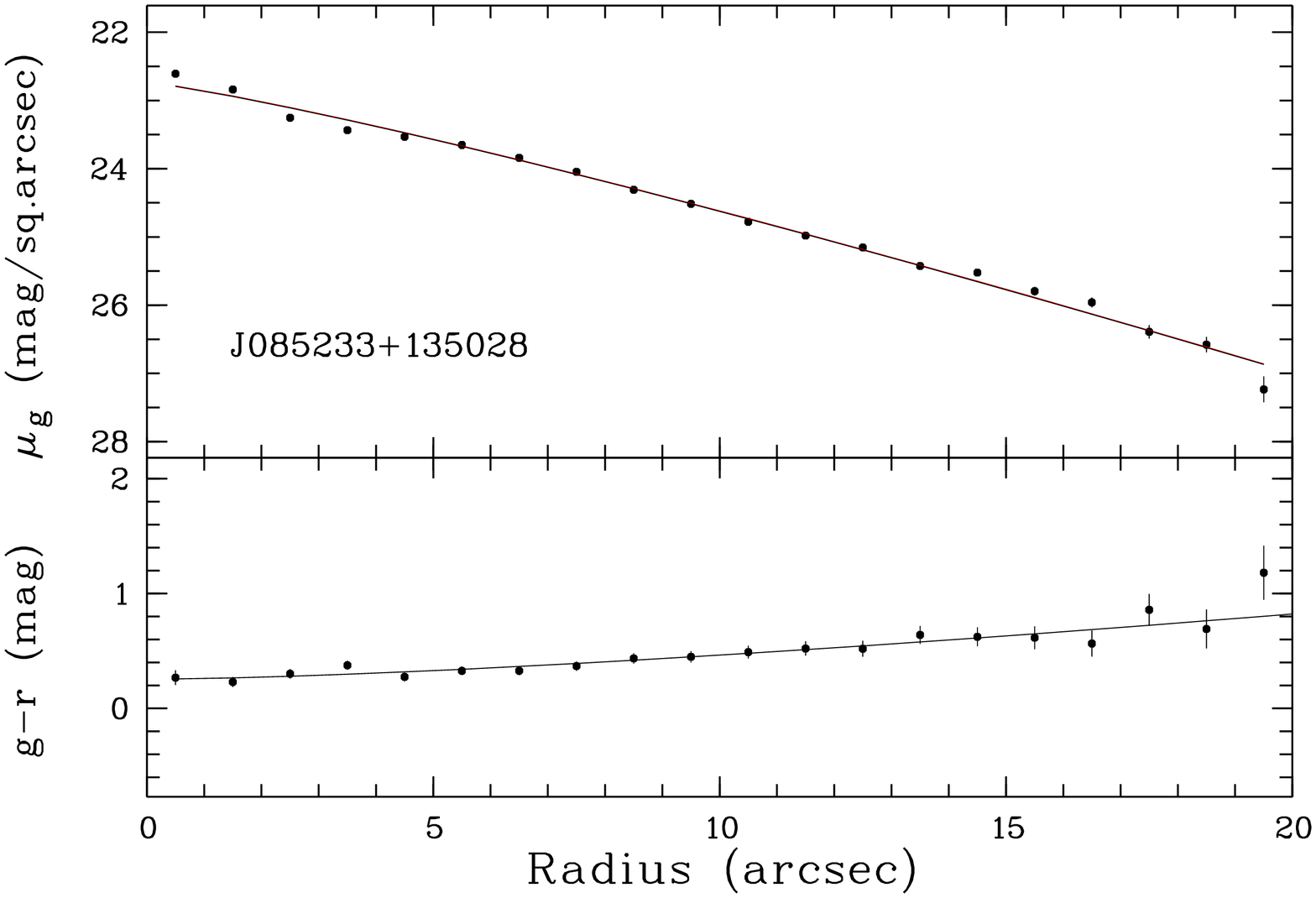}
  \caption{
The profiles of the surface brightness in $g$-filter and the colour $(g-r)$
versus the effective radius for three Lynx-Cancer void LSBDs.
{\bf Top panel:}  J0723+3621.
{\bf middle panel:} J0737+4724.
{\bf bottom panel:} J0852+1350.
}
	\label{fig:SB_profile}
 \end{figure}

\begin{figure}
  \centering
 \includegraphics[angle=-90,width=8.0cm, clip=]{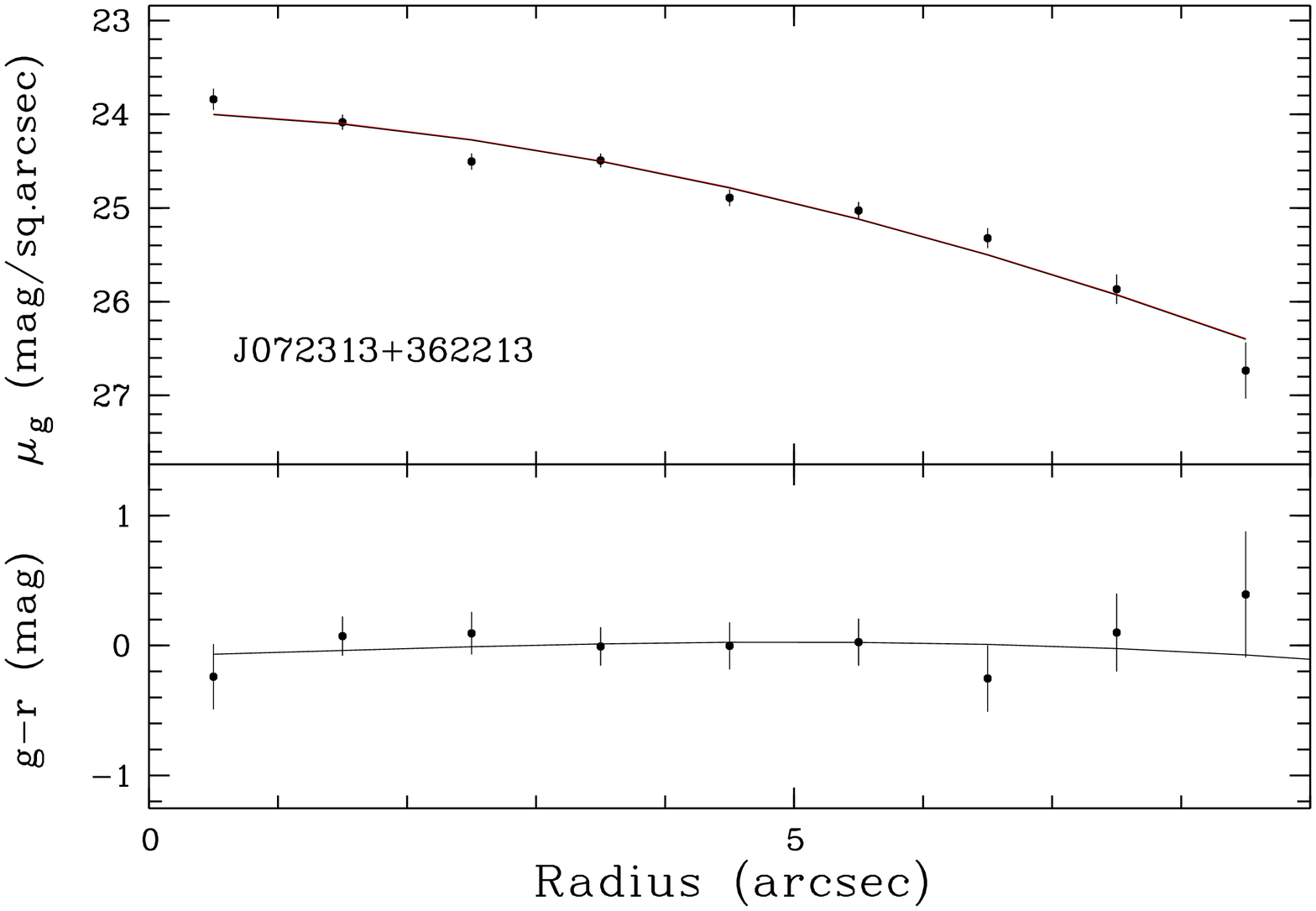}
 \includegraphics[angle=-90,width=8.0cm, clip=]{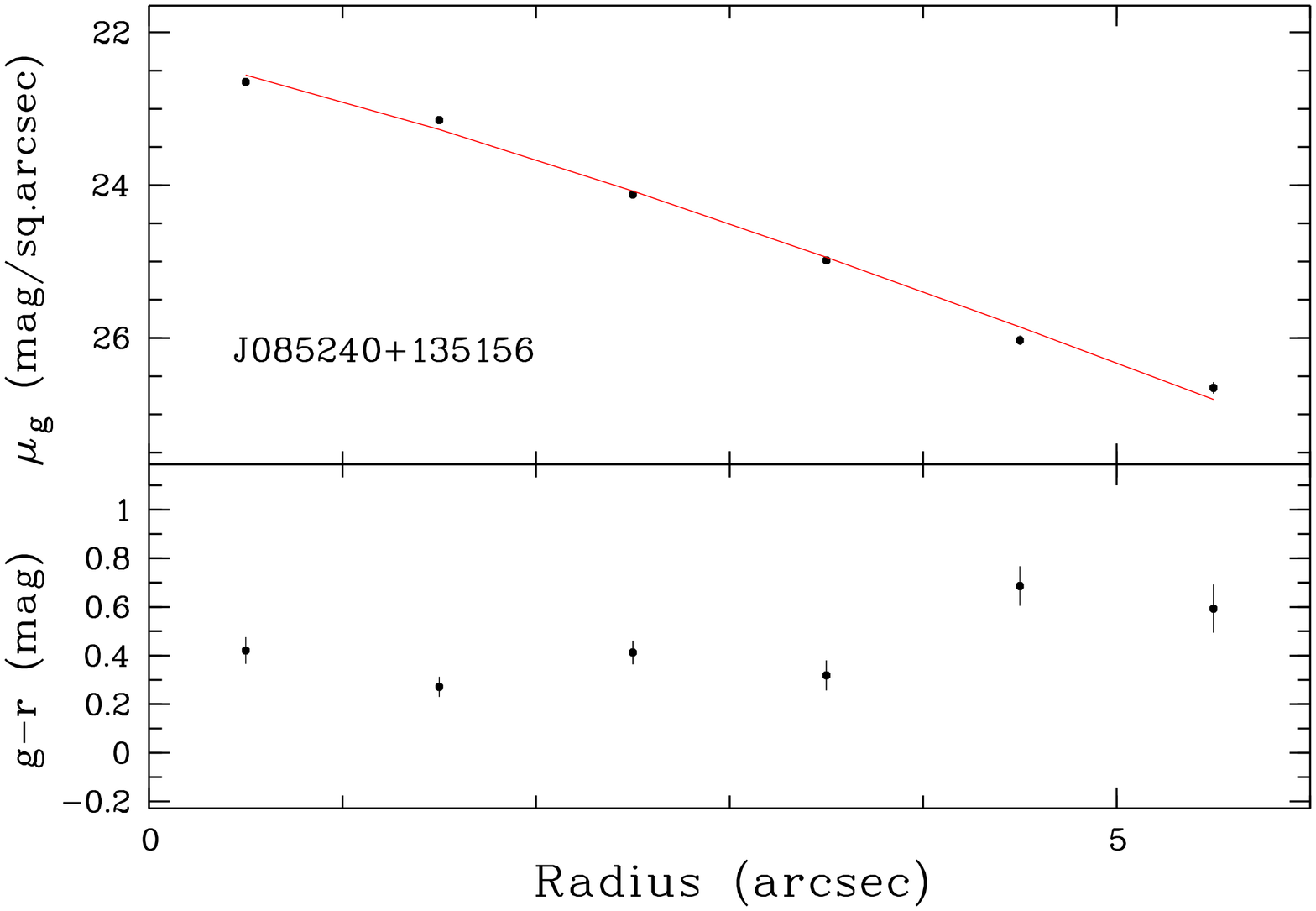}
  \caption{
The profile of surface brightness in $g$-filter and the colour $(g-r)$
versus the  effective radius for two faint companions of the Lynx-Cancer
void LSBDs.
{\bf top panel:} J0723+3622 - a LSBD companion of J0723+3621.
{\bf bottom panel:} J0852+1351 - a dwarf companion of J0852+1350.
}
	\label{fig:SB_profile_comp}
 \end{figure}

In the bottom of Table~\ref{tab:photo} we summarize parameters of the Sersic
profile fitting: the observed central SB in $g$ and $r$ filters, the observed
central SB in $B$-band $\mu_{\rm 0,B}$ (from the two former parameters, using
the transformation formula of \citet{Lupton05}) and the respective value,
corrected for the foreground Galaxy extinction \citep{Schlegel98} and for
disc inclination -- $\mu_{\rm 0,B,c,i}$. For the majority of these galaxies,
the Sersic index $n$ is close to 1, so their SB profiles are close to purely
exponential. Only for the faintest galaxy J0723+3622 ($M_{\rm B}$ = --11.9)
the SB profile looks significantly different from the exponential one.
We also give the 'disc' scalelengths in $g$ and $r$, and the colours $(u-g)$,
$(g-r)$ and $(r-i)$ of the outer regions of the target galaxies, used to
constrain the ages of the oldest stellar population, as shown in the last
line. The latter were
estimated from the additional photometry on a subset of small round diaphragms
(diameter of $\sim$2\arcsec) in the outer regions with the exception of those
with detectable H$\alpha$-emission.
In Fig.~\ref{fig:SB_profile} and \ref{fig:SB_profile_comp} we show,
respectively,  for three new void LSBDs and two their faint companions, the
representative radial surface brightness (SB) profiles in $g$-filter and
the radial run of $(g-r)$ colour.

One of the goals of our surface photometry was to compare the observed
colours of the outer regions with the PEGASE2 model evolutionary tracks
\citep{pegase2}, in order to obtain the age estimates of the oldest visible
stellar population in a galaxy. As one can see from the $(g-r)$ colour
radial profile in Fig.~\ref{fig:SB_profile}, due to the contribution of
\ion{H}{ii} regions to the light of the outer parts of the galaxy,
it is difficult to estimate from the profile the outer colours and decide
whether there is a colour gradient in the underlying stellar population.
Also, due to rather elongated form of the galaxies, the photometry in
ring diaphragms mixes the light from different parts of the `disc', and thus
will tend to wash-out colour gradients if they are present. Therefore, to
analyse the colours in galaxy outer parts, we employ the
approach used in our analysis of the stellar colours in the galaxies DDO~68
and SDSS J0926+3343 \citep{DDO68_sdss,J0926}.

\begin{figure*}
 \centering
 \includegraphics[angle=-90,width=8.5cm,clip=]{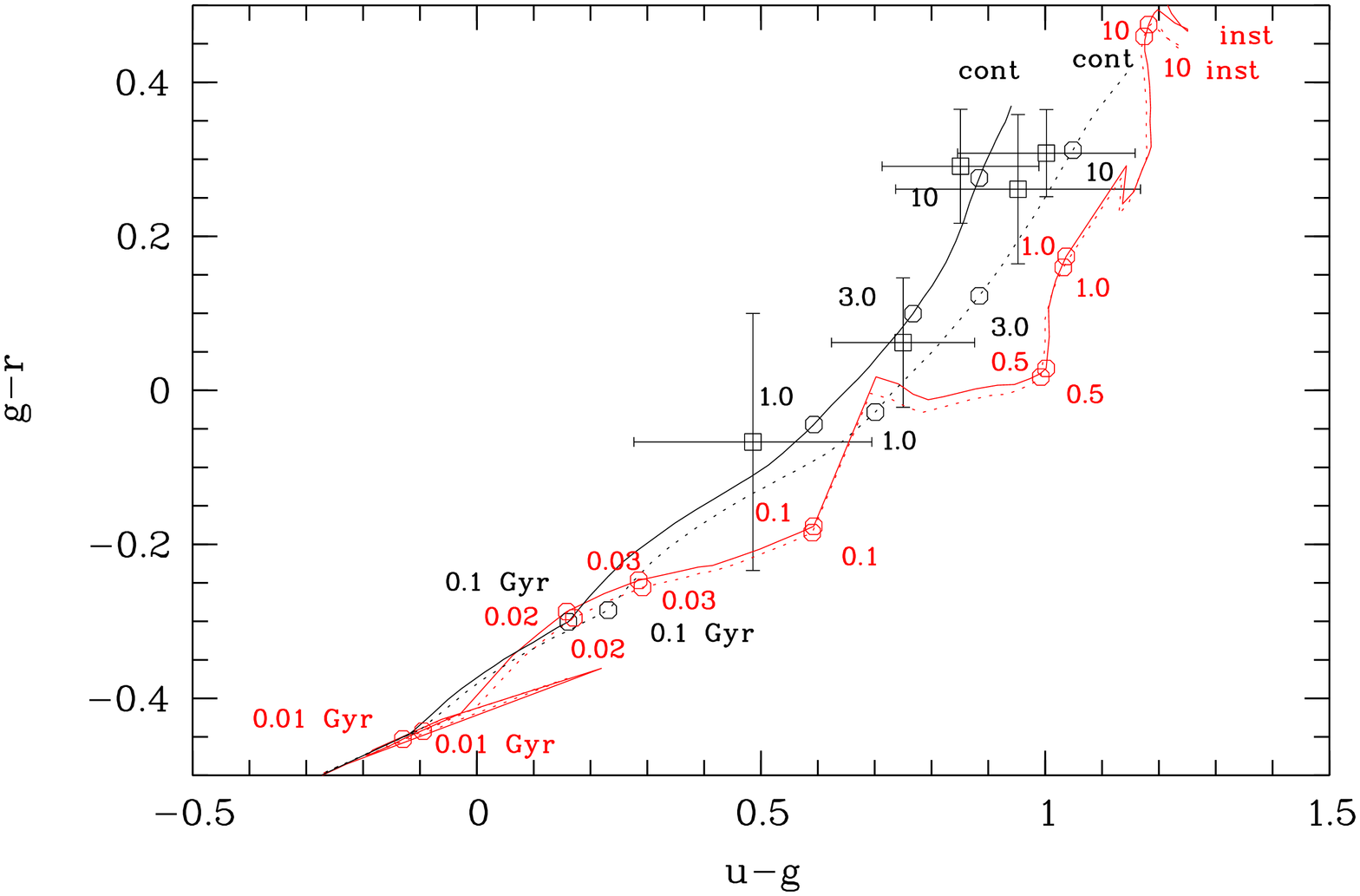}
 \includegraphics[angle=-90,width=8.5cm,clip=]{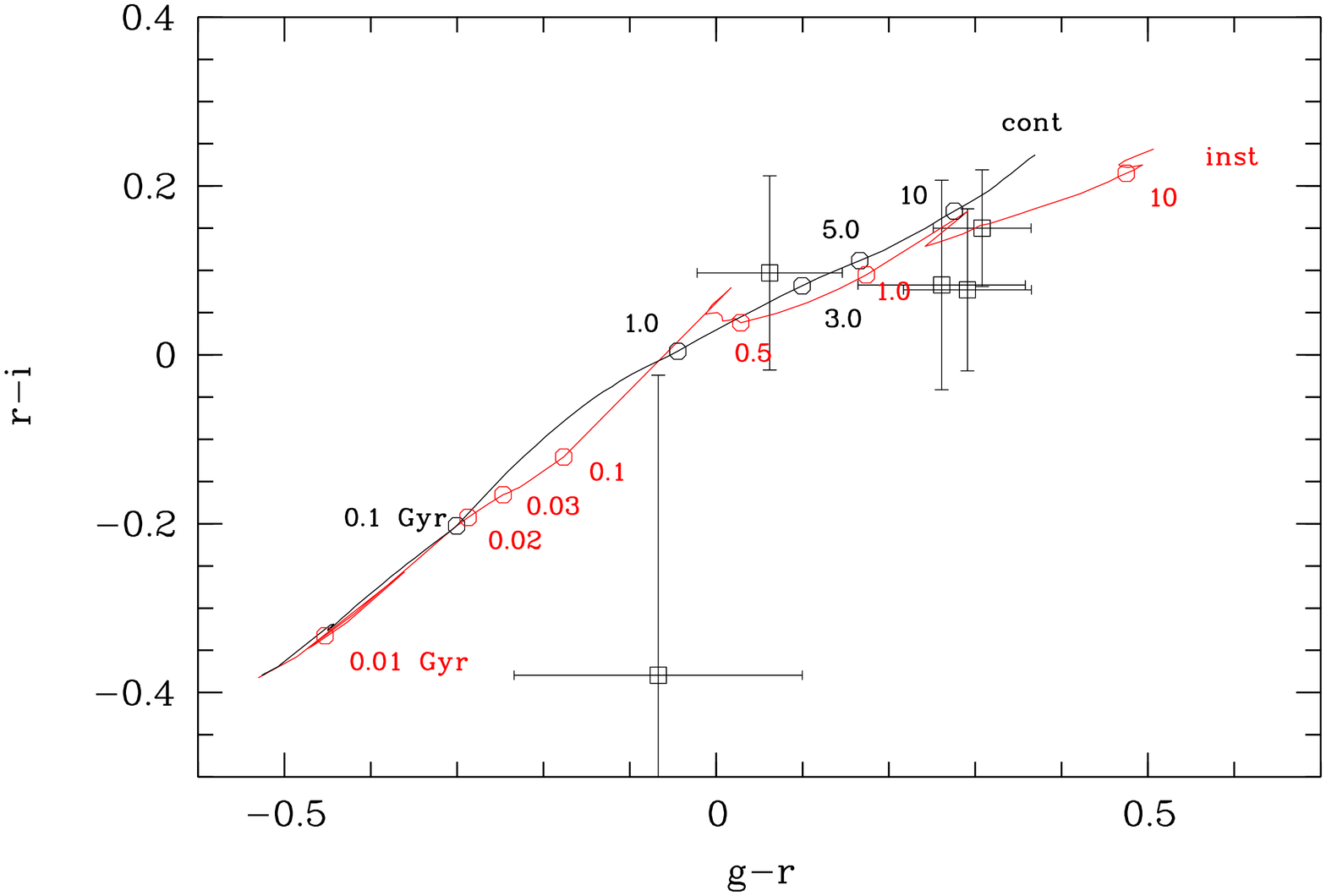}
  \caption{\label{fig:ugri}
Colours of the outer parts of the studied Lynx-Cancer LSBDs, shown by squares
with 1$\sigma$ error bars, are confronted with the PEGASE2 model
evolutionary
tracks for two extreme star formation laws. Tracks for continuous SF with
constant SFR (solid black) and for instantaneous SF (solid red), for the
standard Salpeter IMF and the nearest to the observed metallicities
value of $z$=0.0004, are marked by dots, indicating time in Gyr since the
beginning of a SF episode. The similar tracks for \citet{Kroupa93} IMF are
shown by the dotted lines. See also text for more details.
  {\bf Left panel:} $(g-r)$ vs $(u-g)$. The galaxies with the bluest colours
and 'small' age are respectively, J0723+3622 and J0737+4724.
{\bf Right panel}: Same as in the previous plot, but for $(r-i)$ vs $(g-r)$.
The Kroupa IMF tracks are not shown here since they run very closely to
those for the Salpeter IMF and practically do not add useful constraints.
}
\end{figure*}

In Fig.~\ref{fig:ugri}, the PEGASE2 tracks for metallicity $z$=0.0004 (the
nearest to the observed in these galaxies) are compared with the observed,
extinction corrected colours, $u-g$,$g-r$, $r-i$ of the outer regions with
the negligible nebular emission (as given in the bottom part of
Table~\ref{tab:photo}). The exception is the very compact galaxy J0852+1351
with a nearby confusing background galaxy. For this object we adopted the
integrated colours for the light inside a round diaphragm with a diameter of
$\sim$8\arcsec. The tracks for both: the standard Salpeter and  the
\citet{Kroupa93} IMF with the reduced contribution of the sub-solar mass
stars are shown in the $u-g$,$g-r$ plot. In the $g-r$, $r-i$ plot, the Kroupa
IMF tracks are very close to the Salpeter ones. Due to this degeneracy, we do
not show the former tracks. Also, the tracks for both SF laws run
very close to each other. We show the galaxy colours in this plot
to illustrate the general consistency with those from the $u-g$,$g-r$ plot.
In fact, the photometric systems
($u^{\prime}$,$g^{\prime}$,$r^{\prime}$,$i^{\prime}$,$z^{\prime}$)
used for calculations of the PEGASE2 evolutionary tracks and ($u,g,r,i,z$)
used in the real SDSS observations are slightly different. We applied the
transformation formulae from \citet{Tucker06} in order to correct
theoretical values to the ($u,g,r,i,z$) system.

In the $u-g$,$g-r$ diagram, the colours of the examined regions follow
the model tracks rather well, with a better match for the tracks with
continuous SF law and Kroupa IMF. For three of these LSBDs, the colours
are consistent, within rather large uncertainties, with `cosmological'
ages of $T \gtrsim$(7.5--10)$\pm$(3.5--5)~Gyr. For two the bluest objects -
J0723+3622 and J0737+4724, the colours indicate the ages of $\sim$1~Gyr and
$\sim$2~Gyr, respectively, with the $\pm$1~$\sigma$
confidence intervals of $\sim$0.2--3~Gyr and $\sim$1--3~Gyr.

\begin{table*}
\caption{Photometric parameters of five Lynx-Cancer void LSBDs}
\label{tab:photo}
\begin{tabular}{lccccc} \\ \hline \hline
Parameter                           & J0723+3621         & J0723+3622       & J0737+4724              & J0852+1350                & J0852+1351       \\ \hline
$g_{\rm tot}$                       & 16.66$\pm$0.01     & 19.10$\pm$0.02   & 17.69$\pm$0.01          & 17.05$\pm$0.01            & 19.47$\pm$0.02   \\
$(u-g)_{\rm tot}$                   & 1.23$\pm$0.06      & 0.55$\pm$0.19    & 1.00$\pm$0.08           & 1.08$\pm$0.05             & 0.49$\pm$0.05    \\
$(g-r)_{\rm tot}$                   & 0.40$\pm$0.01      & --0.06$\pm$0.11  & 0.46$\pm$0.03           & 0.49$\pm$0.02             & 0.34$\pm$0.03    \\
$(r-i)_{\rm tot}$                   & 0.01$\pm$0.02      & --0.25$\pm$0.19  & --0.26$\pm$0.04         & 0.05$\pm$0.02             & 0.08$\pm$0.03    \\
$(i-z)_{\rm tot}$                   & --0.12$\pm$0.07    & --               & 0.40$\pm$0.08           & --0.05$\pm$0.02           & ---              \\
$B_{\rm tot}$                       & 17.01$\pm$0.02     & 19.31$\pm$0.03   & 18.06$\pm$0.02          & 17.43$\pm$0.02            & 19.80$\pm$0.03   \\
$R_{\rm g,25}$ ($\arcsec$)          & 13.8               & 5.3              & 8.7                     & 11.6                      & 3.4              \\
$R_{\rm r,25}$ ($\arcsec$)          & 15.7               & 5.5              & 10.4                    & 14.9                      & 3.9              \\
$(b/a)_{\rm 25}$ (SDSS)             & 0.34               & 0.67             & 0.46                    & 0.60                      & 0.64             \\
$R_{\rm g,Hol}$ ($\arcsec$)         & 20.9               & 8.2              & 13.7                    & 18.1                      & 5.0              \\
$R_{\rm r,Hol}$ ($\arcsec$)         & 24.3               & 8.6              & 15.8                    & 20.8                      & 5.7              \\   \hline
$\mu_{0,g}$(mag~arcsec$^{-2}$)      & 22.65$\pm$0.03     & 23.98$\pm$0.13   & 22.88$\pm$0.08          & 22.73$\pm$0.07            & 22.28$\pm$0.11    \\
$\mu_{0,r}$(mag~arcsec$^{-2}$)      & 22.28$\pm$0.08     & 24.06$\pm$0.11   & 22.80$\pm$0.05          & 22.47$\pm$0.05            & 21.79$\pm$0.14    \\
$\mu_{0,B}$(mag~arcsec$^{-2}$)      & 22.99              & 24.13            & 23.19                   & 22.97                     & 22.66             \\
$\mu_{0,B,c,i}$(mag~arcsec$^{-2}$)  & 24.14              & 24.36            & 23.66                   & 23.41                     & 23.18             \\
$n_{\rm g}$(Sersic)                 & 1.14$\pm$0.02      & 1.74$\pm$0.28    & 1.23$\pm$0.06           & 1.17$\pm$0.05             & 1.24$\pm$0.07    \\
$\alpha_{\rm g}$($\arcsec$)         & 7.1                & 5.4              & 5.1                     & 6.2                       & 1.7               \\
$n_{\rm r}$(Sersic)                 & 1.13$\pm$0.06      & 1.94$\pm$0.28    & 1.29$\pm$0.04           & 1.13$\pm$0.04             & 1.05$\pm$0.07     \\
$\alpha_{\rm r}$($\arcsec$)         & 7.1                & 5.6              & 6.3                     & 6.8                       & 1.4               \\
$(u-g)_{\rm outer,c}$               & 0.95$\pm$0.21      & 0.49$\pm$0.22    & 0.75$\pm$0.13           & 1.00$\pm$0.18             & 0.85$\pm$0.14     \\
$(g-r)_{\rm outer,c}$               & 0.26$\pm$0.10      & --0.07$\pm$0.17  & 0.06$\pm$0.08           & 0.31$\pm$0.06             & 0.29$\pm$0.07     \\
$(r-i)_{\rm outer,c}$               & 0.08$\pm$0.12      &-- 0.38$\pm$0.35  & 0.10$\pm$0.12           & 0.15$\pm$0.08             & 0.08$\pm$0.10     \\
$T$(`old' population)               & $\sim$4--13~Gyr    & $\sim$0.1--3~Gyr & $\sim$1--3~Gyr          & $\sim$6.5--14~Gyr         & $\sim$6.5--13.5~Gyr      \\
\hline \hline
\multicolumn{6}{p{12.8cm}}{%
(1) $(b/a)$-- from NED; (2) -- values of $\mu_{0,g}$ and others are corrected
for the Galactic extinction and inclination (see text);
(3) -- colours of the outer parts are corrected for the Galactic extinction. }
\end{tabular}
\end{table*}

\section[]{DISCUSSION}
\label{sec:dis}

\subsection{Main parameters}

In Table~\ref{tab:param} we present the main parameters of the three studied
LSB galaxies. From the total magnitudes in filters $g$ and $r$ (Table
\ref{tab:photo}),  with the transformation equations of \citet{Lupton05}
we derive their values of $B_{\rm tot}$. For the adopted distance moduli
of galaxies J0723+3621, J0737+4724 and J0852+1350, $\mu$=30.97 ($D$=15.6 Mpc),
$\mu$=30.09 ($D$=10.4 Mpc) and  $\mu$=31.82 ($D$=23.1 Mpc), and
the values of the foreground Galactic extinction $A_{\rm B}$=0.23, 0.47 and
0.16~mag, their resulting  absolute magnitudes $M_{\rm B}^0$ are
--14.21, --12.32 and --14.66~mag, respectively.
These values of $M_{\rm B}^0$ correspond to the blue luminosities of
$L_{\rm B}$=6.5$\times$10$^{7}$, 1.32$\times$10$^{7}$ and
11.4$\times$10$^{7}$ $L_{\rm B}$\sunn.  Using the masses $M$(\HI) derived
in the previous section, one obtains the ratio $M$(\HI)/$L_{\rm B}
\sim$3.9, $\sim$1.9 and $\sim$2.6 (in solar
units).

From the \HI-line width at 20\%-level of the maximum intensity,
$W_\mathrm{20}$, one can estimate the maximal rotational velocity, using the
standard approximation, as, e.g., formula (12) from \citet{Tully85}. For
further estimates we account for inclination corrections for the three
galaxies of  1/$\sin(i)$ $\sim$1.04, $\sim$1.12 and $\sim$1.29, respectively.
The inclination angle was adopted according to the standard formula
$cos(i)^2 = (p^2-q^2)/(1-q^2)$, where $p=b/a$ is the observed axial ratio, and
$q$ is the intrinsic axial ratio, adopted to be equal to 0.2.
Then, for the measured line widths $W_\mathrm{20}$=70, 55, and 93~\kms,
respectively, the observed rotation velocities are $V_{\rm rot}$ $\sim$29,
$\sim$21 and $\sim$39~\kms. The inclination-corrected $V_{\rm rot}$ are then
$\sim$30, $\sim$24 and $\sim$50~\kms, for J0723+3621,  J0737+4724 and
J0852+1350, respectively. All these $V_{\rm rot}$ are quite typical of dwarf
galaxies with similar values of $M_{\rm B}$ as traced for the faint dwarf
galaxies \citep[e.g., in FIGGS sample, ][]{Begum08b}.

Having $V_{\rm rot}$ and the characteristic sizes of the studied galaxies,
one can estimate their total (dynamical) masses, which are necessary to
balance the centrifugal force within a certain radius.
The typical radii of \HI\ discs (at the column  density level of
10$^{19}$~atoms~cm$^{-2}$) in dwarf galaxies with $M_{\rm B}$ between
--12.5 and --14.5~mag,
close to that of our LSB dwarfs, are 2.5--3 times larger than the Holmberg
radius \citep[e.g.,][]{Begum08a,Begum08b}. Therefore, we accept that the
\HI\ radii of the studied galaxies are 2.7~$R_{\rm Holm}$. Then,
for galaxies J0723+3621, J0737+4724 and J0852+1350, we have, respectively,
$R_{\rm HI} =$6.57, 2.43, 6.05~kpc.
Finally, from the relationship:
$M_{\rm tot}$($R < R_{\rm HI}$) = $V_{\rm rot}^2~\times$~$R_{\rm HI}$/$G$,
where $G$ is the gravitational constant, one derives the total mass within
$R_{\rm HI}$. They are, respectively,
$M_{\rm tot}$=136$\times$10$^{7}$, 32$\times$10$^{7}$ and
350$\times$10$^{7}$~$M$\sunn.

To estimate a galaxy gas mass, we sum $M$(\HI) and $M$(He)
(assuming the mass-fraction of 0.33 for He) and obtain
(in the same order): $M_{\rm gas}$=38.6$\times$10$^{7}$,
3.43$\times$10$^{7}$ and 35.2$\times$10$^{7}$~$M$\sunn.
These estimates have the interesting implications. The galaxies, as expected,
are dark-matter dominated, but not too much. Taking into  account the
contribution of the stellar mass (which is quite small, see below), the ratio
$M_{\rm tot}/M_{\rm bary}$ is $\sim$7.5--9.1 for J0737+4724 and J0852+1350,
and only $\sim$3.0--3.4 for J0723+3621.
These values (except that of J0723+3621) are similar to those for
two other extreme void dwarfs: DDO~68 and J0926+3343  ($\sim$5.3
and $\sim$12) \citep[][and Table~\ref{tab:param}]{DDO68,J0926}
and do not differ much from the universal value 1/$f_{\rm bar}$ =
$\Omega_{\rm matter}$/$\Omega_{\rm bary}$ = 0.27/0.0469$\sim$5.8, derived
from the most updated WMAP7 data \citep{Jarosik10}. For the galaxy
J0723+3621, due to the rather approximate decomposition of its complex
\HI\ profile, we may have underestimated the parameter $W_\mathrm{20}$
and the
related values of $V_{\rm rot}$ and $M_{\rm tot}$. In any case, for a better
estimate of $M_{\rm tot}$ in these LSBDs, one needs their resolved \HI\
kinematics.

To estimate the stellar mass of the studied galaxies, we follow the recipe
of the stellar $M/L$ ratio ($\Upsilon$) from the recent paper by
\citet{Zibetti09}. To compare the gas mass-fractions in our galaxies with
those in LSBGs from the other studies, we also obtain the stellar masses
via the $\Upsilon$ from the frequently used recipes in \citet{Bell03}.
We notice that both these recipes use the initial mass functions (IMF)
with the significant deficit of the subsolar-mass stars with respect to the
standard Salpeter IMF. Hence, the resulting values of $\Upsilon$ are
noticeably smaller.

We take from Table \ref{tab:photo} the total $g$-magnitudes and the colours
$(g-i)^{0}$ of the integrated light, corrected for the foreground Galactic
extinction and calculate $\Upsilon$ on a linear fit using the
coefficients given in Table B1 of \citet{Zibetti09}. We use this $\Upsilon$
as the most robust value according to \citet{Zibetti09}. Similar values
of $\Upsilon$ are obtained using to the coefficients in Table~7  of
\citet{Bell03}. The range of the derived estimates of stellar masses with
these two prescripts for each of the five LSBDs is given in
Table~\ref{tab:param}. Despite both recipes having approximately the same IMF
for subsolar mass stars, the resulting values of
$\Upsilon$ derived via the recipe from \citet{Zibetti09} are smaller by
a factor of $\sim$4 to $\sim$9, depending on the colour.
As \citet{Zibetti09} explain this difference, this is mainly
related not to the different IMF, but to the different assumptions about
SF history in terms of ages and bursts.

We also provide the related estimates of $f_{\rm gas}$, the ratio of
$M_{\rm gas}/M_{\rm bary}$.  For comparison, we also present similar
estimates of $M_{*}$ and $f_{\rm gas}$ for the other Lynx-Cancer void
galaxies added to Table~\ref{tab:param}.

\subsection{Galaxy masses versus model predictions}
\label{sec:masses-models}

It is interesting to compare the derived values of
$M_{\rm tot}$/$M_{\rm bary}$ and the estimated $V_{\rm circ}$ for the five
mentioned above extreme void dwarfs with the predictions from the numerical
simulations by \citet{Hoeft06,Hoeft10}, who found a significant baryon
deficiency for objects with $V_{\rm circ} \lesssim$40~\kms\ due to the UV
radiation heating of the infalling baryon gas. One should account
for the difference in $M_{\rm tot}$ definition in the model DM halos.
The latter corresponds to the virial mass of a DM halo $M_{\rm vir}$, defined
at the radius $R_{\rm vir}$. For dwarfs, the radius at which $V_{\rm rot}$
is close to the flat part of the rotation curve, is a factor of 5--10 smaller
than $R_{\rm vir}$ \citep[see, e.g.,][]{Trujillo10}. The related mass of a
DM halo is therefore much higher than the directly measured at the last point
of the observed rotation curve. With this difference in mind, the
correspondence of our estimates to the results of simulations for the ratio
M$_{\rm tot}$/M$_{\rm bary}$ versus $V_{\rm circ}$, shown in Fig.~2h of
\citet{Hoeft10}, can be considered acceptable at the zero approximation for
the galaxies with $V_{\rm rot} \lesssim$ 30~\kms. However, the resolved
\HI-mapping and modelling of the mass distribution are necessary to make
the comparison more reasonable.
As for the two galaxies with $V_{\rm rot}$ = 50--55~\kms, their observed
baryonic mass-fractions appear too small. Namely, after correction of the
measured M$_{\rm tot}$ to that of the DM halo, they
fall much lower with respect to the predicted baryon mass-fraction.

Another option of the examining the dynamical properties of the studied void
LSBDs is to confront them with the results of recent state-of-art models. In
particular, \citet{Trujillo10}, using the $\Lambda$CDM cosmological
simulations of the very large volume, present the Baryonic Tully-Fisher (BTF)
relation
(their Fig.~11) for $V_{\rm circ}$ measured at $R = 10$ kpc,
which well matches the parameters of galaxies in the wide range of
$V_{\rm circ}$=60-400~\kms. Of course, the data of \HI-mapping are
prerequisite for a more reliable comparison with the model predictions since
one cannot be confident on the real extent of the rotation curve and its
inclination correction (due to the possible difference in inclination of the
optical body and its \HI-disc).
However, the baryon masses of the two fastest rotators -- J0852+1350 and
DDO~68 -- are not much different to the model-predicted values. For DDO~68
we use
the GMRT \HI-maps from \citet{Ekta08}. But this object is likely the result
of a recent merger. The smaller void LSBDs fall outside the modelled rotation
velocity range, but all of them have much too large baryonic masses as
compared with naive extrapolation of the theoretical curve. This part of the
BTF curve, is defined, according to \citet{Trujillo10}, mostly by the
`observed' luminosity function of SDSS galaxies. One of the possible
solutions of the apparent problem is that for the void LSBD galaxies with
$V_{\rm circ}$ = 20--30~\kms\ their baryonic mass-fractions are 5--10 times
larger than the more typical SDSS galaxies. One needs larger dwarf galaxy
samples in the wide range of the environment density to probe the parameter
space which affects the galaxy baryonic mass-fraction.

\begin{figure*}
 \centering
 \includegraphics[angle=-90,width=8.5cm,clip=]{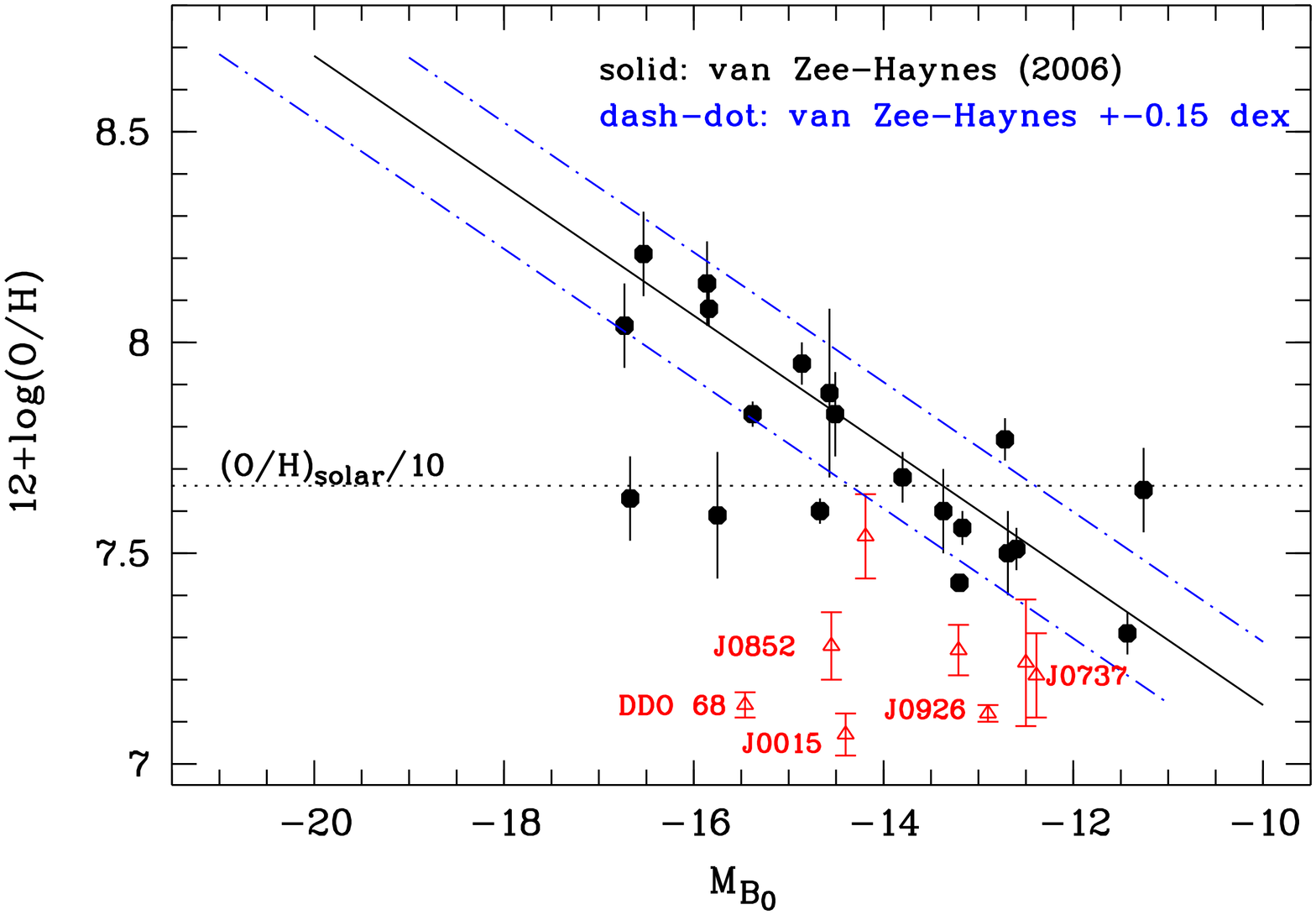}
 \includegraphics[angle=-90,width=8.5cm,clip=]{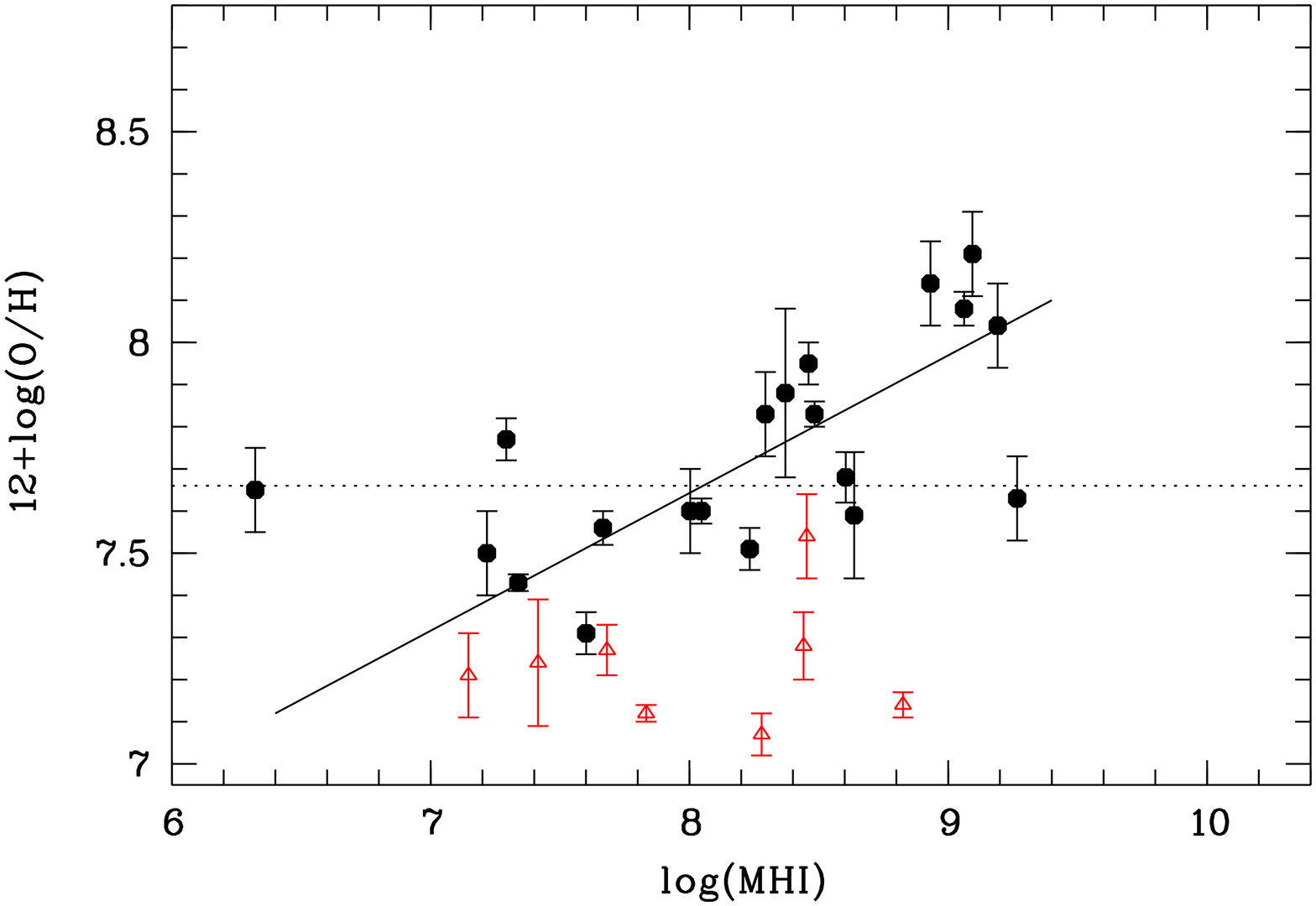}
  \caption{\label{fig:ZvsMB}
{\bf left panel:} O/H values of the discussed 'unevolved' Lynx-Cancer LSBDs
[and SDSS J0744+2508 and J0812+4836 from Paper~II and SDSS J0015+0105 from
\citet{Guseva09}, which resides in the Eridanus void (Pustilnik et al.
2011, in preparation)  versus their $M_{\rm B}$
(red triangles) in comparison to the fit for the Local Volume late-type
dwarfs from \citet{vZee06a} (solid black line) with two dash-dot lines below
and above that showing the rms of that fit of 0.15 dex. For comparison,
we show 14 LSBGs from the Local Volume and its surroundings from
\citet{vZee06a} (black octagons with the error bars) and 6 objects
from the literature (see text).
{\bf right panel:} The similar plot, but for O/H versus the galaxy \HI-mass.
Solid line shows approximate fit for 19 LSBGs (excluding the left-most dot)
of \citet{vZee06a}. \HI-data for XMD LSBDs are from this paper and the cited
articles, for J0015+0104 -- from the paper in preparation by
Pustilnik et al. (2011).
}
\end{figure*}

\subsection{Faint companion LSBDs}
\label{sec:companions}

The incidence of dwarf pairs and interactions in voids was already noticed
in earlier studies \citep[e.g., by][]{HIvoids,SAO0822} and recently by
\citet{VGS}. Here we add the information on two more dwarf pairs in the
Lynx-Cancer void.

At $\sim$2.6$\arcmin$ NEE from J0723+3621, approximately along the
direction of its major axis,  there is a faint very blue LSB galaxy SDSS
J072313.46+362213.0 (hereafter, J0721+3622), with the total size
of $\sim$10\arcsec ($\sim$1~kpc), $g$=19.10, $g-r$= --0.06 and $r-i$= --0.25.
From the BTA long-slit spectrum described in Sec.~\ref{BTA}, with
both galaxies on the slit, we derived on H$\alpha$-line the radial velocity
of J0721+3622, which is  $\sim$90~\kms\ larger than that of J0723+3621.
This implies that the galaxies are in a rather high-velocity encounter.
The higher spectral resolution spectrum (with FWHM=5.5~\AA)  of J0723+3621
is insufficient for a good quality rotation curve in H$\alpha$. However,
this permits a determination of the direction of rotation and to obtain
a rough estimate of its amplitude: the eastern edge is receding and
the full amplitude of rotation is $\sim$40~\kms\ at the radial distances
$\pm$17\arcsec. The respective value of $V_{\rm rot} \sim$20~\kms\ is by
a factor $\sim$1.4 lower than that derived from the \HI\ profile (see
Table~\ref{tab:param}). This is quite natural since the H$\alpha$ estimate
corresponds to the internal region where the full extent is several times
smaller than that for $V_{\rm rot}$ based on the \HI\ profile.

From the analysis of the two \HI\ profiles in Fig.~\ref{fig:HI},
corresponding to the NRT pointings to two both components of this pair,
the first approximation of the \HI\ gas distribution can be inferred.
This consideration should account for the signal drop due to the beam off-set.
The lower velocity peak, corresponding to that of J0723+3621, drops by
a factor of $\sim$5 at the antenna beam pointing to J0723+3622. However,
the second peak, containing  approximately the same amount of \HI-flux, does
not change for this pointing as one could expect. This suggests that the
\HI\ gas responsible for the emission in this velocity range is situated
mid-way between the two galaxies and probably represents a strong tidal
bridge. Since the total luminosity of J0723+3622 in $B$-band is only
$\sim$1/8 of that for J0723+3621, the strong tidal effect
in this collision poses the question on the relative total masses of both
galaxies. The orientation of the mutual orbital motion and of the rotation
vector for J0723+3621, determined on the long-slit spectrum in this paper,
indicates that this galaxy experiences a prograde encounter. The latter
usually infers a strong tidal effect. But still, the strong tidal
bridge needs a significant mass for the disturber. The \HI\ mapping of the
system should help to better estimate the dynamical mass of J0723+3622.
Besides, the N-body simulations of the system can better constrain the
minimal total mass of this very faint, but quite a massive LSBD.

Similarly, at $\sim$2$\arcmin$ NE from the galaxy J0852+1350 ($\sim$13~kpc in
projection), approximately along its minor axis, there is a faint ($\sim$9
times less luminous) blue Irr galaxy, SDSS J0852+1351, with a total size of
$\sim$8\arcsec\ ($\sim$1~kpc), $g$=19.47, $g-r$=0.34 and $r-i$=0.08.
As described in Sec.~\ref{obsJ0852}, its radial velocity, obtained on the BTA
spectrum is $\sim$30~\kms\ higher than that of its larger companion.
A somewhat disturbed morphology of J0852+1350, with a circular type
feature at the SE edge, can be considered as an indication on the substantial
effect of interaction between the two objects.

These two pairs can be the analogs of other binary/interacting systems of
the very gas-rich metal-poor galaxies such as the well known SBS~0335--052E,W
\citep{VLA,Ekta09,Moiseev10}  and the Lynx-Cancer void pair HS~0822+3542 and
SAO~0822+3545 \citep{SAO0822,Chengalur06}. It is worth noting that in such
very metal-poor and gas-rich pairs the optical luminosity ratio
may not provide a useful indicator of the baryon or the total mass ratio.
In particular, the $B$-band light ratio for the E and W galaxies in
SBS~0335--052 system is $\sim$7:1 \citep{SBS0335_BTA}, while the \HI-mass
ratio is $\sim$0.8:1, and the total mass ratio is of $\sim$1 \citep{Ekta09}.
For the other mentioned pair we see rather the opposite
situation. The LSBD galaxy SAO 0822+3545 is a factor $\sim$1.4 more luminous
and a factor of $\sim$3 is more massive in \HI. In such galaxies \HI-mass
is a good proxy of the baryon mass, since their stellar mass contributes to
baryons no more than 5--20\%.
Therefore, to understand whether these two void LSB pairs are  minor
or major encounters, it is crucial to obtain their \HI-maps.

\subsection{Unusual LSBDs in Lynx-Cancer void versus more typical LSBDs}

In the previous sections we described several new gas-rich Lynx-Cancer void
LSBDs with unusual properties: the extremely low O/H and/or blue outer colours
which imply a relatively recent main episode of star formation. In
Table~\ref{tab:param} we included three additional similar objects in this
void
identified in our earlier papers - SAO 0822+3545, SDSS J0926+3343 and DDO~68.
It is reasonable to ask, whether these `extreme/unusual' void LSBDs really
differ from the more common LSBDs.

One of the options is to check how they are similar to other known LSBDs in
respect of various relationships involving O/H. Also, since they are very
gas-rich, it is reasonable to compare them  to LSBDs
with the large value of $M$(\HI)/$L_{\rm B}$. Many such dwarf galaxies are
collected in the FIGGS sample \citep{Begum08b}.
Unfortunately, it is not an easy task - to obtain the reliable O/H estimates
in LSBGs, since their H{\sc ii} regions are rather faint. On the other hand,
for many late-type dwarfs in the Local Volume, the reliable O/H data are
available. However, the data in their surface photometry
with the estimates of the central SB of the underlying LSB disc are absent.
The samples with known O/H and the SB data are poorly correlated
and to separate LSBD galaxies with known O/H is a problem.
Therefore, we use for comparison the relationship 'O/H versus $M_{\rm B}$'
from \citet{vZee06a}. There are two more complementary samples from the
Local Volume, those of \citet{vZee06b} and \citet{Lee03}. They result in
similar relationships 'O/H versus $M_{\rm B}$',  but with somewhat
higher scattering for the fit lines.

For comparison with other correlations between optical and \HI\ properties,
one can use the \HI-selected galaxies from the Equatorial Survey (ES)
based on HIPASS
\citep{Garcia09,West09}. These authors, in particular, found in their sample
of 195 galaxies a fraction of the smallest objects assigned to the type of
"inchoate" galaxies which are blue dwarf LSB gas-rich objects apparently
similar to part of our void sample objects. Their inchoate objects are
not outliers. From the relationships between the whole sample
galaxy parameters, they comprise probably the low-mass tail of the broad
parameter distributions of the \HI\ identified galaxies. Therefore, it is
of interest to understand how the parameters other than O/H for our
`extreme' void LSBDs are similar to those of the "inchoate" type.

In the left panel of Fig.~\ref{fig:ZvsMB} we show the five `extreme'
Lynx-Cancer LSBDs (red triangles) from Table \ref{tab:param} in the diagram
`O/H vs $M_{\rm B}$'.  The solid line shows the general relationship for
late-type dwarfs from the Local Volume and its surroundings, as derived by
\citet{vZee06a}. As explained in Paper~II, we applied the small correction
to their regression line to convert their original O/H to the new O/H scale
[in which our O/H data are given \citep{Izotov06}].
The filled dots show 13 LSBGs from
\citet{vZee06a} sample and seven  additional LSBDs from the literature.
The latter include And~IV \citep{Begum08b,AndIV}, SDSS J1201+0211 and
J1215+5223 \citep{Kniazev03,NRT_07}, ESO489-56, ESO577-27 and ESO146-14
\citep{Ronnback95,MG96,MF96,NRT_07}, DDO~154 (NED).
The red triangles show two more very metal-poor void
galaxies SDSS J0744+2508 and J0812+4836 from Paper~II and \citet{IT07}
and a new LSBD SDSS J0015+0104 with 12+$\log$(O/H)=7.07 \citep{Guseva09},
which is very isolated from luminous galaxies and resides deeply in the
Eridanus void.

The majority of LSBDs from the `control' sample appear within the region
of $\pm$0.15~dex from the `standard' fit line of \citet{vZee06a}. This
region is shown in the plot as limited by the two dashed lines, parallel to
the `standard' line. Indeed, as \citet{vZee06a} demonstrate, the LSB members
of their late-type galaxy sample do not show any systematic deviation
from their `standard' fit.
However, the two southern metal-poor LSBDs ESO577-27 and ESO146-14 from
\citet{Ronnback95} fall significantly below this border ($\sim$0.3-0.4~dex)
and this might hint on the systematic shift of LSB metallicities relative
to their high SB counterparts. A newly identified by \citet{Mattsson11}
very metal-poor LSB galaxy ESO~546-G34 (12+$\log$(O/H) $\sim$7.4~dex),
probably resembles the discussed here LSB dwarfs, but its environment is
not studied.

The statistics of LSBDs with well known O/H
and the good photometry is not yet large enough to check the impact of
LSB galaxies on to the slope and the zero-point of `O/H vs $M_{\rm B}$'
relationship. As shown in
Paper~II, the effect of the global environment on to the dwarf galaxy
metallicity does exist. Therefore, to probe the effect of LSB in galaxies,
one needs to separate a LSB sample in a more typical global environment.
Four of the eight `extreme' void galaxies fall significantly below the
lower dashed line. In this sense they are indeed XMD (eXtremely
Metal-Deficient) as defined by \citet{EC10}. That is their values of
O/H are 3-7 times lower than those for typical similar galaxies residing
in denser environments.  This is indicative of their unusual evolutionary
history.

In the right panel we show a similar plot, `O/H vs $M$(\HI)'
where the hydrogen mass $M$(\HI) is used as a surrogate of the total
baryonic mass. Here the scatter of the control LSBD sample is larger.
But again, the three Lynx-Cancer void galaxies with the lowest metallicities
and SDSS J0015+0104 show O/H significantly smaller than their analogs
from denser environment regions with the same gas mass.

\begin{figure*}
 \centering
 \includegraphics[angle=-90,width=8.5cm,clip=]{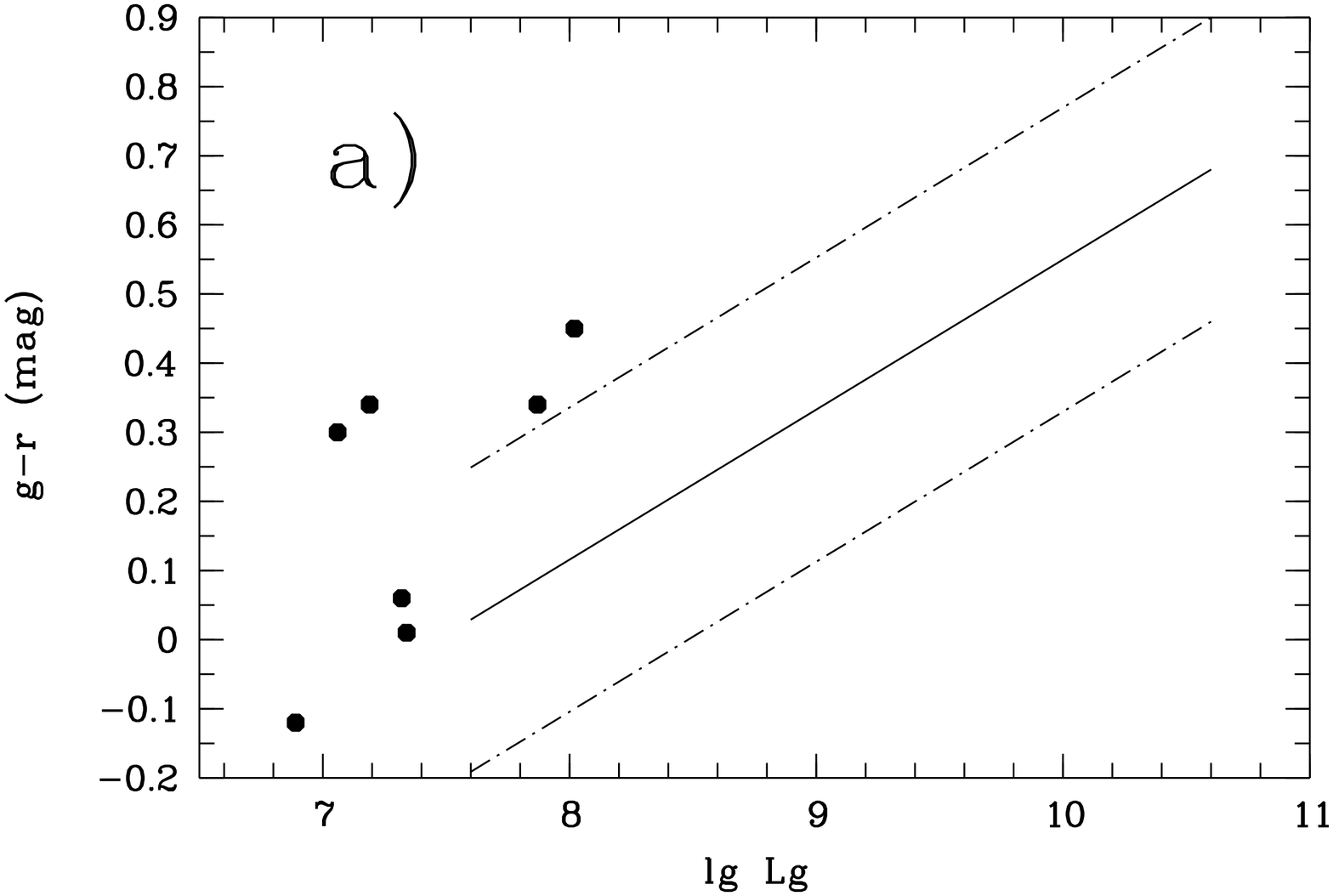}
 \includegraphics[angle=-90,width=8.5cm,clip=]{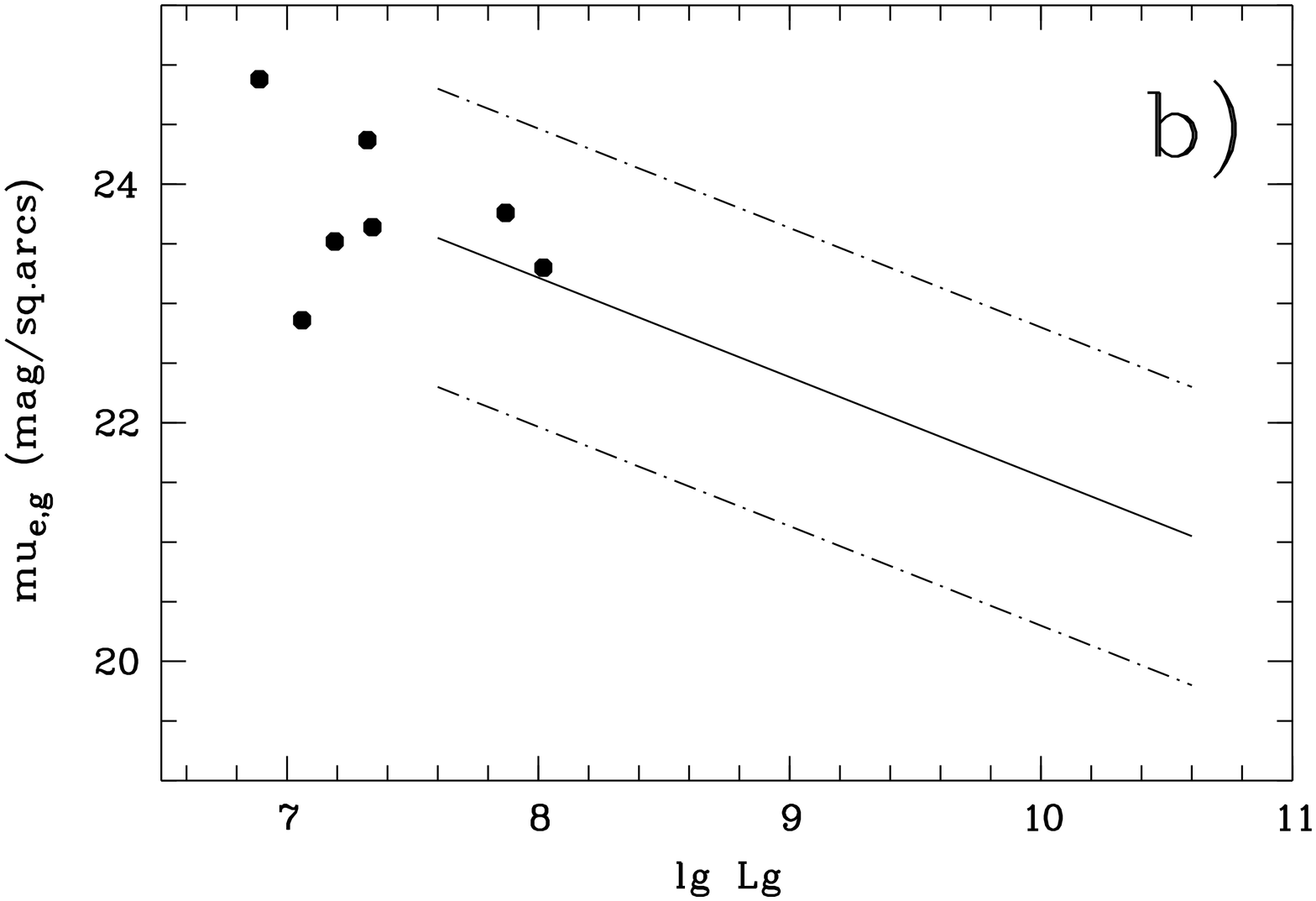}
 \includegraphics[angle=-90,width=8.5cm,clip=]{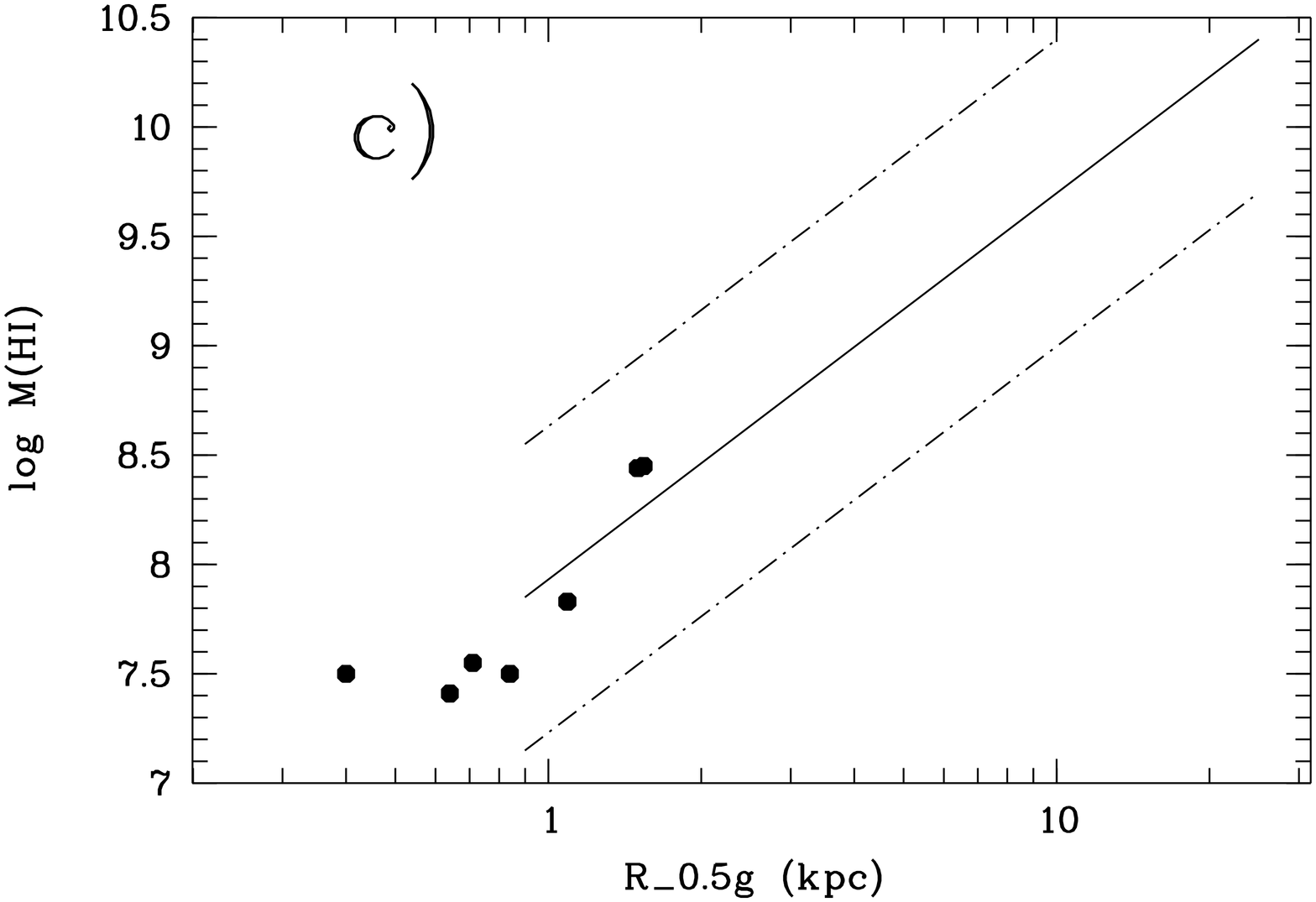}
 \includegraphics[angle=-90,width=8.5cm,clip=]{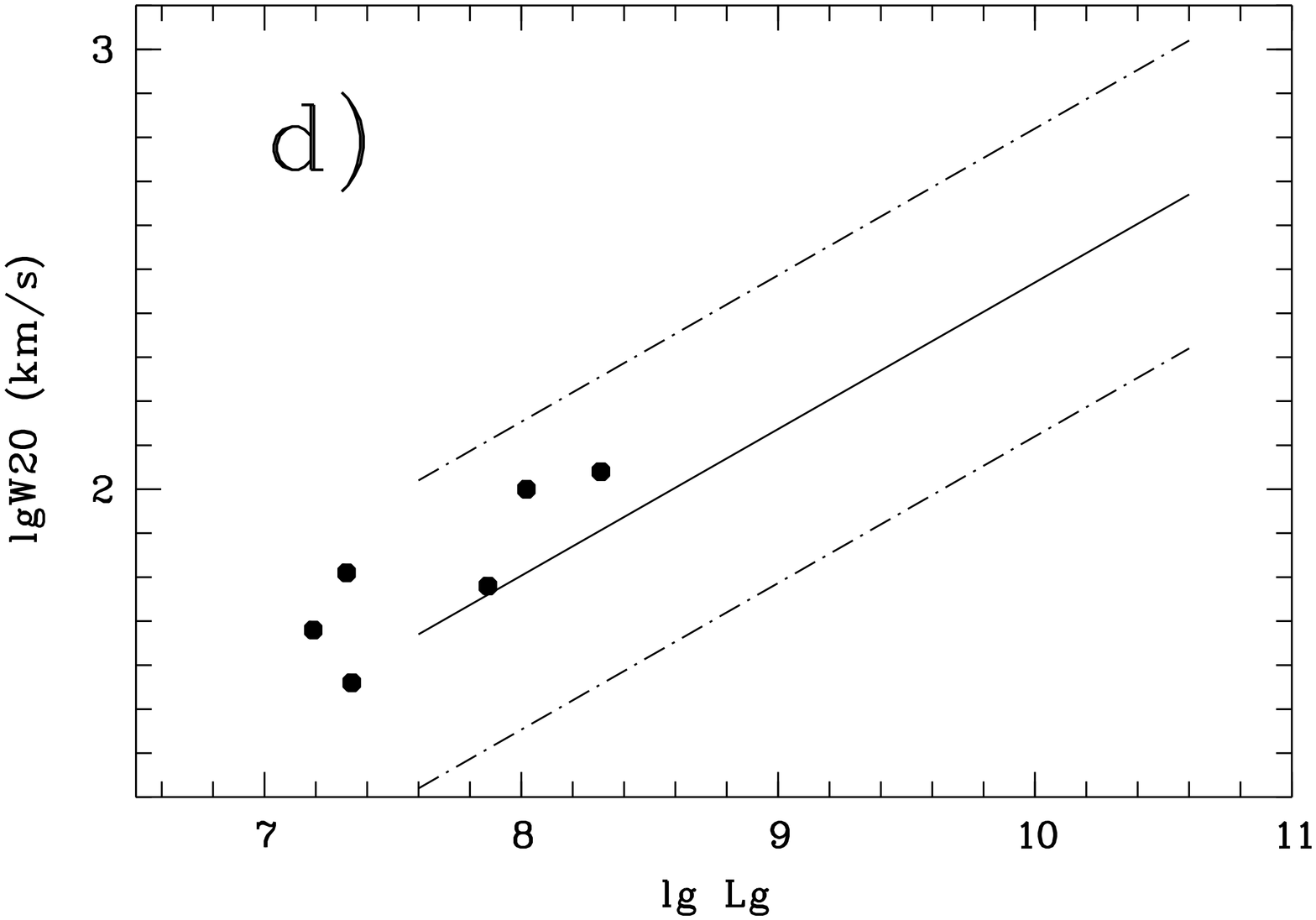}
  \caption{\label{fig:ES_compar}  Comparison of the Lynx-Cancer void
extreme LSBDs from this paper (filled circles) with galaxies from the
ES sample of \citet{Garcia09}.
The solid lines in all plots show linear fits to ES galaxy data, while
the dotted lines limit the range including $\sim$90\% of all ES data points.
{\bf a)} the total $(g-r)^{0}$  colours vs $\lg L_{\rm g}$ ($g$-filter
luminosity in solar units). {\bf b)} the surface brightness/luminosity
relation $\mu_{\rm eff}$(g) vs lg(L$_{\rm g}$.
{\bf c)} the $M$(HI) (in solar units) vs $R_{\mathrm 50,g}$ (effective
radius in kpc) relation. {\bf d)} the Tully-Fisher relation $W_{\mathrm 20}$
vs $\lg L_{\rm g}$.
}
\end{figure*}

We have checked how the parameters of these extreme void LSBDs match to the
correlations
between the main parameters of the ES sample of \HI\ identified galaxies from
\citet{Garcia09}  on their plots $g-r$ vs $L_{\rm g}$, $\mu_{\rm eff}$(g) vs
$L_{\rm g}$, $M$(\HI) vs $R_{\mathrm 50}$(g), $W_{\mathrm 20}$(\HI) vs
$L_{\rm g}$.  In Fig.~\ref{fig:ES_compar}a,b,c,d  we show respectively the
sketches of their Figures 15, 16, 17 and 19 with the data for our void
galaxies overlaid. For two faint companion LSBDs without \HI-flux estimates
we adopted conditionally the same $M$(\HI)/$L$ ratios as for their brighter
counterparts.
The luminosities and hydrogen masses as well as the
\HI\ profile widths of the considered void LSBDs appear on the lower edges
of the respective ES sample distribution or even outside these ranges.
But in none of these plots the void galaxies in the question show
the deviations larger than the typical scatter of ES sample galaxies from
their general trends. This suggests that the global properties of the void
LSBDs are similar to the those of the dimmest, `inchoate' part of the
ES sample. Currently neither metallicities,
nor the global environment are known for these inchoate ES galaxies.
In principle, they can be the full analogs of the void lowest metallicity
LSBDs. In the case of their O/H are higher and closer to the
`standard' `O/H vs $M_{\rm B}$' relationship, they  can be just more
evolutionary advanced cousins of rather rare `extreme' void LSBDs.
The study of  both parameters, O/H and global environment of the inchoate
ES galaxies will help to understand the possible relation between
the unusual Lynx-Cancer void galaxies and the specific subsamples of
galaxies  separated in large `unbiased' surveys.

Completing the issue of unusual LSBDs residing in this void, we
emphasize their high spatial concentration. The full volume of this void with
diameter $D=8$~Mpc is $\sim$1.5 times larger than that for the Local Volume
defined in \citet{CNG} as a sphere with $D=7$~Mpc, containing $\sim$450
galaxies with known distances. Most of them are situated in a flattened
Local Sheet \citep{Tully08} with a typical density exceeding the global
value.
For the significant fraction of the Local Volume galaxies, the parameter
O/H is available. Only for two of them - Leo~A and UGCA~292, their values
of O/H
appear to be as low as 12+$\log$(O/H) $\sim$7.3. But both of them match the
\citet{vZee06a} `O/H vs $M_{\rm B}$' relationship, so they just follow the
general dependence. The volume of the Lynx-Cancer void relative to that of a
large sphere with radius of 26~Mpc, in which the Lynx-Cancer void is
immersed, comprises only $\sim$5\%. There are at least four extremely
metal-poor LSBDs in the void (including DDO~68 and SDSS J0926+3343 at
the mutual distance of only 1.6~Mpc). We can count only one similar
galaxy I~Zw~18 for the large sphere. Its real distance is still debated, but
the analysis of its environment shows that I~Zw~18 is also well isolated,
having no luminous neighbours closer than 2~Mpc.

Since no complete study of O/H in galaxies of the LV and the Local
Supercluster exists, one can not exclude the role of some selection effects
in this drastic enhancement of concentration of the `extreme' metal-poor
LSBDs in the Lynx-Cancer void. However, the overall effect of this
"selection" does not look to be significant. Nevertheless, the absence of
more or less complete database for O/H in
the LV galaxies is an important caveat for the comparative study of galaxy
evolution in various environments.

The increased density of such unusual objects in the void region is
certainly not by chance. The most metal-poor galaxies prefer the
low galaxy-density regions. However, even in voids their fraction is small.
The void environment can be favourable for the retarded dwarf galaxy
formation and their slower evolution due to significantly reduced effect
of galaxy encounters and their induced SF. Thus the `inchoate' galaxies
could be a significant fraction of void population. On the other hand,
the large fraction of currently detected `extreme' void dwarfs appear to
experience a recent stronger or weaker interaction resulted in the extra SF
and the enhanced intensities of emission lines. This eased their discovery,
determining their optical redshifts and their identification as the
void galaxies.

How large a fraction of the dormant gas-rich galaxies is undisclosed
in the void, remains an open question. The well-known interacting/merging
pair of the most metal-poor dwarf galaxies
SBS~0335--052E,W \citep[see new data in][]{Ekta09,Izotov09} is also situated
near the border of a large void \citep{Peebles01}. The existence of the
sizable \HI\ cloud population in `nearby' voids, with the baryonic masses
comparable to those of dwarf galaxies, visible through their Ly-$\alpha$
absorption \citep[][ and references therein]{Manning02,Manning03},
also hints on the unevolved state of the significant
fraction of baryons in voids. The more advanced analysis of the Lynx-Cancer
void dwarf galaxy census and the summary of their properties will be
presented elsewhere.

\begin{table*}
\caption{Main parameters of new LSBDs and of three other known extreme
dwarfs in the void}
\label{tab:param}
\footnotesize{
\begin{tabular}{lcccccccc} \\ \hline \hline
Parameter                           & J0723+3621           & J0723+3622              & J0737+4724              & SAO~0822             & J0852+1350        &  J0852+1351    & J0926+3343        & DDO~68              \\ \hline
A$_{\rm B}$ (from NED)              & 0.23                 & 0.23                    & 0.47                    & 0.20                   & 0.16              &  0.16          & 0.08              & 0.08                \\
B$_{\rm tot}$                       & 17.01$\pm$0.03       & 19.31$\pm$0.03          & 18.06$\pm$0.03          & 17.56                  & 17.43             &  19.80         & 17.34             & 14.60               \\
V$_{\rm hel}$(H{\sc i})(\kms)       & 888$\pm$2$^{(3)}$    & 954$\pm$3$^{(3)}$       & 473$\pm$2$^{(3)}$       & 742$\pm$2              & 1507$\pm$3        &  1537$\pm$22   & 536$\pm$2         & 502$\pm$2           \\
V$_{\rm LG}$(H{\sc i})(\kms)        & 885$\pm$2$^{(3)}$    & 951$\pm$3$^{(3)}$       & 521$\pm$2$^{(3)}$       & 711$\pm$2              & 1360$\pm$3        &  1390$\pm$22   & 488$\pm$2         & 428$\pm$2           \\
Distance (Mpc)                      & 15.6                 & 15.6                    & 10.4                    & 13.5                   & 23.1              &  23.1          & 10.7              & 9.9                 \\
M$_{\rm B}^0$ $^{(4)}$              &  --14.19             &  --11.89                &  --12.50                & --13.30                & --14.55           & --12.18        & --12.90           & --15.46             \\
Opt. size (\arcsec)$^{5}$           & 44$\times$15         & 12$\times$9.5           & 21.6$\times$10.4        & 28.2$\times$15.5       & 24.8$\times$15.4  & 10.0$\times$6.4& 35.8$\times$9.9   & 103$\times$38       \\
Opt. size (kpc)                     & 3.2$\times$1.1       & 0.88$\times$0.69        & 1.12$\times$0.54        & 1.85$\times$1.0        & 2.78$\times$1.72  & 1.12$\times$0.7& 0.93$\times$0.26  & 4.94$\times$1.82    \\
$\mu_{\rm B,c,i}^0$ $^{(4)}$        & 24.14                & 24.36                   & 23.66                   & 23.40$^{(8)}$          & 23.41             & 23.18          & 25.4$^{(9)}$      & 23.3$^{(11)}$        \\
12+$\log$(O/H)                      & 7.5:                 &  --                     & 7.24$\pm$0.20           & ....                   & 7.28$\pm$0.08     & ---            & 7.12$\pm$0.02     & 7.14$\pm$0.03       \\
\HI\ int.flux$^{(6)}$               & 5.05$\pm$0.18        &  --                     & 1.01$\pm$0.08$^{(2,3)}$ & 0.83$\pm$0.08          & 2.10$\pm$0.15     & ---            & 2.54$\pm$0.07     & 28.9$\pm$3.0        \\
W$_\mathrm{50}$ (km s$^{-1}$)       & 46$\pm$3$^{(2,3)}$   &  --                     & 39$\pm$4$^{(2,3)}$      & 33.8$^{(9)}$           & 76$\pm$5          & ---            & 47.4$^{(9)}$      & 81.4$^{(12)}$       \\
W$_\mathrm{20}$ (km s$^{-1}$)       & 70$\pm$5$^{(2,3)}$   &  --                     & 55$\pm$6$^{(2,3)}$      & ...                    & 93$\pm$8          & ---            & 80.5$^{(9)}$      & 103.0$^{(12)}$      \\
V$_\mathrm{rot}$ (H{\sc i})(\kms)   & 29.8$^{(2,3)}$       &  --                     & 23.8$^{(2,3)}$          & 18$^{(9)}$             & 49.8$^{(2,3)}$    & ---            & 32$^{(9)}$        & 55$^{(12)}$         \\
M(H{\sc i}) (10$^{7} M_{\odot}$)    & 28.4                 &  --                     & 2.60                    & 3.57                   & 27.6              & ---            & 6.8               & 66.8                \\
M$_{\rm dyn}$ (10$^{7} M_{\odot}$)  & 136                  &  --                     & 32$^{(2)}$              & 13.9                   & 350$^{(2)}$        & ---           & 124               & 490                 \\
M(H{\sc i})/L$_{\rm B}$$^{(7)}$     & 3.9$^{(2)}$          &  --                     & 1.9$^{(2)}$             & 1.1                    & 2.6$^{(2)}$       & ---            & 3.0               & 2.9                 \\
M$_{*}$ (10$^{7} M_{\odot}$)        & 1.37, 6.13           &  0.02, 0.18             & 0.10, 0.66              & 0.20, 1.17             & 3.21, 12.05       & 0.23, 1.00     & 0.14, 0.91        & 1.66, 10.0          \\
f$_{\rm gas}$                       & 0.96, 0.86           &  --                     & 0.97, 0.84              & 0.96, 0.80             & 0.92, 0.75         & ---           & 0.98, 0.91        & 0.98, 0.90          \\
T(main pop.)                        & $\sim$5-13~Gyr$^{(2)}$ & $\sim$0.1--3~Gyr      & $\sim$1--3~Gyr          & 1--2~Gyr               & $\sim$6--13~Gyr   & $\sim$6--13~Gyr& 1--3~Gyr          & $\sim$1~Gyr         \\
\hline \hline
\multicolumn{8}{p{16.6cm}}{%
(1) -- from NED; (2) -- derived in this paper; (3) -- derived from NRT \HI\
profile;  (4) -- corrected for the Galactic extinction A$_{\rm B}$ and
inclination ($\mu_{\rm B,c,i}^0$); (5) -- full size $a \times b$ at
$\mu_{\rm B}=$25\fm0~arcsec$^{-2}$;
(6) -- in units of Jy~\kms; (7) -- in solar units;
(8) \citet{SAO0822}; (9) - \citet{J0926};  (10) - \citet{IT07};(11) -
\citet{DDO68_sdss}; (12) - \citet{Ekta08}. }
\end{tabular}   }
\end{table*}

\subsection{Summary}
Summarising the results and discussion above, we draw the following
conclusions:

\begin{enumerate}
\item
Three LSB dwarf galaxies SDSS J0723+3621, J0737+4724 and J0852+1350
($\mu_{\rm B,c,i}^0$=24.14, 23.66, and 23.41~mag~arcsec$^{-2}$,
respectively) situated in the nearby Lynx-Cancer void appear to be the new
representatives of very metal-poor and/or gas-rich void objects. The
oxygen abundances of J0737+4724 and J0852+1350, estimated via the
semi-empirical method,  correspond to 12+$\log$(O/H)=$\sim$7.24 and 7.28
dex, respectively.
\item
All three new void LSBDs (and three other similar void objects) are DM
dominated, with the estimated mass ratios of M$_{\rm tot}$/M$_{\rm bary}
\sim$3.4, $\sim$9 and $\sim$8.8, respectively. The values of this
parameter for four LSBDs with $V_{\rm rot} =$18--32~\kms\ are roughly
consistent with those predicted in the model of \citet{Hoeft10}  with
the suppressed gas accretion to small DM halos due to the UV-heating.
On the other hand, the baryonic
mass of the same four void LSBDs appears too large in respect of
extrapolated theoretical BTF relation from \citet{Trujillo10}, which well
describes the `common' SDSS galaxies. These two findings do not match each
other. This issue certainly needs a deeper analysis and comparison with
the most updated simulations.
\item
In the close surroundings ($\sim$12--13 kpc in projection) of the two
studied void LSBDs J0723+3621 and J0852+1350, we found the faint blue
LSBDs SDSS J0723+3622 and J0852+1351, with $\mu_{\rm B,c,i}^0$=24.36 and
23.18~mag~arcsec$^{-2}$.  Their relative velocities are  +89~\kms\ and
+30~\kms, respectively. The parameters of encounters for the two systems
correspond to a prograde and a polar collision, respectively. The \HI\ data
indicate significant gas exchange in the former pair.
\item
The $(u-g)$, $(g-r)$ and $(r-i)$ colours in the outer regions of three of the
five studied void galaxies derived from the photometry of their SDSS images,
appear rather typical of other galaxies. They correspond to the PEGASE2
model track for the evolving stellar population with the continuous SF for
ages of $T \sim$5--13 Gyr (both Salpeter of Kroupa IMF). For J0737+4724 and
the faint LSBD companion J0723+3622, the colours of the outer regions appear
bluer, corresponding to the same tracks, but for ages of $\sim$1--2 Gyr.
The combination of parameters O/H, the gas mass-fraction and the colours of
the outer parts of galaxies  indicates the slower rate of chemical
evolution for all these LSBDs, and suggests a significant delay in the
beginning of the main SF episode for galaxies J0723+3622 and J0737+4724.
\item
The discovery of new extremely metal-poor and/or `unevolved' dwarf galaxies
(J0723+3622, J0737+4724, J0852+1350) significantly extends the list of
several such objects previously known as representatives of the Lynx-Cancer
void (namely, DDO~68, SDSS~J0926+3343, SDSS~J0744+2508, SAO~0822+3545,
SDSS~J0812+4836). The very existence of such a numerous group gives an
evidence for the significant concentration of unevolved objects within
a relatively  small cell of the nearby Universe.
The void occupies $\lesssim$5~\% of the volume of the sphere with
R = 26 kpc, to which it belongs and contains almost all such known objects.
This supports the idea on the specific role of voids in providing the most
suitable conditions for the survival of unevolved galaxies on the
cosmological timescale.
\end{enumerate}

\section*{Acknowledgements}

The main part of the paper was prepared during the stay of SAP in Paris,
supported by the grant from the University of Paris-Diderot. SAP is also
thankful to Observatoire de Paris (GEPI) for hospitality. The work on this
project was supported through RFBR grant No.~10-02-92650 to SAP, and through
RFBR grant No.~11-02-00261 to SAP and ALT.  SAP and ALT acknowledge also the
support of this work in the frame of the Russian Federal Innovation Program
(contract No. 14.740.11.0901). AYK acknowledges support from the
National Research Foundation of South Africa. SAP, ALT and AYK acknowledge
the BTA Time Allocation Committee for the continuous support of this project
at the SAO 6-m telescope.
The authors thank A.~Burenkov, D.~Makarov and R.~Uklein for help with BTA
observations. SAP is grateful to A.~Klypin for useful consultations on the
results of recent N-body simulations from \citet{Trujillo10}.
SAP and JMM acknowledge the NRT Time Allocation Committee for the support
of this program in 2009 and 2010.
The authors acknowledge the spectral and photometric data and the related
information available in the SDSS database used for this study.
The Sloan Digital Sky Survey (SDSS) is a joint project of the University of
Chicago, Fermilab, the Institute for Advanced Study, the Japan Participation
Group, the Johns Hopkins University, the Max-Planck-Institute for Astronomy
(MPIA), the Max-Planck-Institute for Astrophysics (MPA), New Mexico State
University, Princeton University, the United States Naval Observatory, and
the University of Washington. Apache Point Observatory, site of the SDSS
telescopes, is operated by the Astrophysical Research Consortium (ARC).
This research has made use of the NASA/IPAC Extragalactic
Database (NED), which is operated by the Jet Propulsion Laboratory,
California Institute of Technology, under contract with the National
Aeronautics and Space Administration.


\bsp

\label{lastpage}

\end{document}